\documentclass[superscriptaddress,showpacs,twocolumn,aps,pre,10pt]{revtex4-1}
\usepackage{natbib}
\usepackage{graphicx,dcolumn,bm,bbm,amsmath,amssymb,color}

\newcommand{\eeq}{ \end{equation} }
\newcommand{\beq}{ \begin{equation} }
\newcommand{\bea}{\begin{eqnarray}}
\newcommand{\eea}{\end{eqnarray}}
\newcommand{\ga}{\alpha}
\newcommand{\gb}{\beta}

\newcommand{\bhu}{ \hat{\bf u} }

\newcommand{\br}{ {\bf r} }

\begin{document}
\title{Mechanisms of carrier transport induced by a microswimmer bath}

\author{Andreas Kaiser}
\email{kaiser@thphy.uni-duesseldorf.de}
\affiliation{Institut f\"ur Theoretische Physik II: Weiche Materie,
Heinrich-Heine-Universit\"at D\"{u}sseldorf,
Universit{\"a}tsstra{\ss}e 1, D-40225 D\"{u}sseldorf, Germany}
\author{Andrey Sokolov}
\affiliation{Materials Science Division, Argonne National Laboratory, 9700 South Cass Avenue, Argonne, Illinois 60439, USA}
\author{Igor S. Aranson}
\affiliation{Materials Science Division, Argonne National Laboratory, 9700 South Cass Avenue, Argonne, Illinois 60439, USA}
\author{Hartmut L\"{o}wen}
\affiliation{Institut f\"ur Theoretische Physik II: Weiche Materie,
Heinrich-Heine-Universit\"at D\"{u}sseldorf,
Universit{\"a}tsstra{\ss}e 1, D-40225 D\"{u}sseldorf,
Germany}

\date{\today}


\begin{abstract}
Recently, it was found that a  wedgelike microparticle (referred to as "carrier")
which is only allowed to translate but not to rotate exhibits a directed translational motion
along the wedge cusp if it is exposed to a bath of microswimmers. Here we model this effect in detail by resolving the 
microswimmers explicitly  using interaction models with different degrees of mutual alignment. 
Using computer simulations we study the impact of these interactions
on the  transport efficiency of V-shaped carrier. We show that the transport mechanisms itself strongly depends on the
degree of alignment embodied in the modelling of the individual swimmer dynamics. For weak alignment, 
optimal carrier transport occurs in the turbulent microswimmer state and is induced by swirl depletion inside
the carrier. For strong aligning interactions, optimal transport occurs already
in the dilute regime and is mediated by a polar cloud of swimmers in the carrier wake pushing the wedge-particle
 forward. We  also demonstrate that
the optimal shape of the carrier leading to maximal transport speed depends on the kind of interaction model used.
\end{abstract}

\maketitle

\section{Introduction}
\label{sec:intro}
The collective properties of active fluids have been studied intensively in the last years
~\cite{Romanczuk2012,Cates_2012,Marchetti_Rev,aranson_ufn}. Examples of such active systems can be found in quite different 
areas of nature ranging from bacteria~\cite{2007SoEtAl,Shenoy_PNAS,Schmidt2008,Poon},
alga~\cite{SwimmingAlga,CRein2010,BMF_PRL14}, spermatozoa~\cite{2005Riedel_Science,Friedrich,Woolley}, 
animals~\cite{Vicsek_Report2012} -- like birds~\cite{BirdsPNAS}, fish~\cite{FishPNAS} or insects~\cite{locusts,Ants} 
-- and even human beings~\cite{Schadschneider,Helbing,Silverberg:13}. All of these systems can be categorized 
as {\it living} active systems. Additionally, there are a broad class of {\it artificial} realizations, based on
various propulsion mechanisms like pure body rotation \cite{ScrewSwimmer}, propulsion by attached flagella ~\cite{Bibette,Ignacio,MagnetoSperm} or various chemically induced 
mechanisms like self-diffusiophoresis~\cite{PaxtonJACS2004,SyntheticRod2005,Howse_2007,BocquetPRL2010,KapralJCP2013,SanchezJACS2011} or self-thermophoresis~\cite{Sano_PRL2010,Bechinger_SM11,BtH_L_part}.

All of these swimmers are known to form spatiotemporal active states like swarming ~\cite{2011Herminghaus,yang-gompper,Ginelli,2012SwinEtAl} and turbulence (swirling) ~\cite{Sokolov_PRL12,2008Saint_Shelley,PNAS,Li_PRE,aranson2013aquatic,gompper2013,CollectiveSpheres,ISA-PNAS14,WiolandPNAS2014}. 
Most of these patterns can be obtained by a simple modeling based on excluded volume effects of effective 
anisotropic objects~\cite{PNAS,WensinkJPCM}. However, the actual particle collisions and the degree of alignment
 are supposed to play an 
important role~\cite{ActiveRollers,Huepe_PRL}. A mutual alignment of colliding swimmers provides a mechanism of swarming
~\cite{vicsek_prl}, as observed for a lot of artificial active systems~\cite{Huepe_PRL,08kudrolli,2007NaRaMe,SenPNAS13}.
On the other hand experiments have shown a swim-off effect of two bacteria after the collisions ~\cite{2007SoEtAl,aranson2007,drescher2011fluid}.

In the last years, in experiment as well as in simulations, active fluids have been considered 
in the presence of passive objects or obstacles.
Fixed boundaries have been shown to
guide active particles~\cite{DenissenkoPNAS,TrappingSperms} and accumulate them
~\cite{Wensink2008,MenzelLaningConf,Lee13Wall,ElgetiGompper13,HaganConfinement}.
This effect has been used to rectify the motion of swimmers
~\cite{Reichhardt_PRL2008,CatesTrapping,StarkPRE13Rectification,RectifyBrazil,Chaikin2007},
and building sorting~\cite{hulme,Reichhardt,MarconiPRE,JulichSortingMicrochannel}
as well as trapping devices for microswimmers~\cite{TrappingSperms,Kaiser_PRL,TrappingOGS}.
The motion of passive particles submersed in complex systems~\cite{TracerHolmPRL}, like active fluids, 
has been studied as well. Starting with
simple small spherical~\cite{TracerGoldstein,TracerClement,TracerDiffusion2000} and curved
~\cite{MalloryArxiv} tracer particles to large deformable chains~\cite{KaiserPolymer}, showing a regime of 
ballistic motion as known for active particles themselves~\cite{Howse_2007,BtH2011,Kaiser_PRE2013Janus}.

Recently, it has been shown that energy can be extracted from active fluids and 
biomolecular motors~\cite{HessReview}. Asymmetric cogwheels submersed in an 
active bath spontaneously rotate~\cite{DiLeonardoPRL,SokolovPNAS,LeonardoPNAS,GearRobots}. Moreover, it has been demonstrated
recently that wedgelike microparticles can be transported in the highly dilute
~\cite{AngelaniCARGO} and  in the turbulent state~\cite{KaiserSokolov_2014} of the active bath.

Here, we focus on the modelling of the dynamics of a wedgelike microparticle 
(referred to in the following as "carrier") when it is exposed to a bath of microswimmers.
These V-shape particles  can be  fabricated by photolithography~\cite{BtH_L_part,SokolovPNAS,Clement_PRL13} 
and can be submersed into a bath of biological or artificial microswimmers.
By using an external field, the carrier is only allowed to translate but not to rotate. In this paper, we resolve the 
microswimmers explicitly  using different interaction models with different degrees of alignment after
a binary collision. We study the impact of these interactions
on the  transport efficiency of the carrier in detail and show that the transport mechanisms itself strongly depends on the
degree of alignment embodied in the modelling of the individual swimmer dynamics.  
In the modelling so far~\cite{KaiserSokolov_2014},
the maximal carrier speed occurred in the turbulent state and caused by  swirl depletion.
Here we show that for strongly aligning interactions the picture is different:
 an even higher transport speed can be achieved in a dilute active 
fluid which is induced by a smectic ordered cluster in the wake  of the carrier.
In contrast to earlier work~\cite{AngelaniCARGO}, here a full swarm has developed 
to push the carrier in an efficient way.
We also compute the shape of the carrier maximizing the transport speed and 
show that it depends on the kind of interaction model used.

This paper is organized as follows: First we specify our modelling in Sec.~\ref{Sec:model} before we 
study and compare the collective behavior in Sec.~\ref{Sec:Bulk}. In Sec.~\ref{Sec:Transport} we study the transport 
efficiency of a wedgelike carrier  and work out the underlying mechanisms, which are directly linked to
swimmer-swimmer interactions. Furthermore, we compute the optimal shape of the carrier which leads to 
the maximum transport speed. Finally, we conclude in Sec.~\ref{conc}.


\section{Model}
\label{Sec:model}

\begin{figure}
\centering
\resizebox{0.9\columnwidth}{!}{\includegraphics{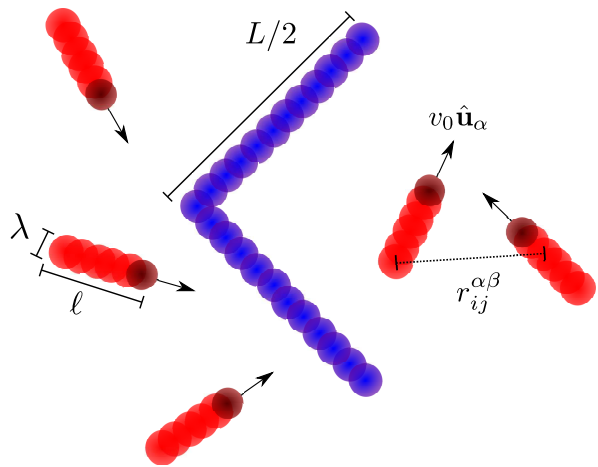}}
\caption{Sketch of the system of self-propelled rods with aspect ratio $\ell / \lambda$ and an effective self-propulsion 
velocity $v_0$ directed along the main rod axis $\bhu$. The single six Yukawa segments are shown by red circles -- a larger 
interaction prefactor for the first segment of each rod is indicated by darker color. A wedgelike carrier with a short contour
length $L$ is shown by blue circles.}
\label{fig:Sketch}   
\end{figure}

We model the active bath in two spatial dimensions by considering  $N$ rodlike self-propelled particles
with center-of-mass positions $\br_{\alpha}$ and orientations $\bhu_{\alpha}$ ($\alpha = 1,...,N$), 
using a possible effective body shape asymmetry 
analogous to Ref.~\cite{KaiserSokolov_2014} in the absence of noise and hydrodynamic interactions. 
Therefore, each rod of length $\ell$ and width $\lambda$ is discretized 
into $n=6$ spherical segments equidistantly positioned along the main rod axis $\bhu=(\cos \varphi, \sin \varphi)$.
The aspect ratio of the swimmers is fixed to $\ell / \lambda = 5$ according to a previous work regarding the 
explicit realization {\it Bacillus subtilis}. Between the segments of different rods a repulsive Yukawa potential 
is imposed~\cite{Kirchhoff1996}. The resulting total pair potential of a rod pair ${\ga,\gb}$ is given 
by $U_{\ga \gb} = \sum_{i=1}^{n}\sum_{j=1}^{n}{U_i U_j} \exp [-r_{ij}^{\ga \gb} / \lambda]/r_{ij}^{\ga \gb}$
where $\lambda$ is the screening length defining the particle diameter, and 
$r_{ij}^{\alpha \beta} = |{\bf r}_{i}^{\ga} - {\bf r}_{j}^{\gb}|$ the distance between segment 
$i$ of rod $\alpha$ and $j$ of rod $\beta$ ($\alpha \neq \beta $). 
The effective body shape of the rods can be tuned by the interaction prefactor of the first segment of each rod
with respect to the others, see Fig.~\ref{fig:Sketch}. This quantity will be given by the ratio 
$U^{\ast} = U_1^2 / U_j^2$ ($j=2\ldots n$), where $U^{\ast}=1$ refers to a symmetric rod. Any overlap of particles 
is avoided by imposing a large interaction strength $U_j^2 = 2.5 F_0 \ell$.
The shape asymmetry allows us to controll the degree of alignment during a binary rod-rod collision
~\cite{Wensink2014}. Here we study {\it  two situations\/} in detail: first of all, the symmetric case where  $U^{\ast}=1$, 
to realize an nematic alignment like in Refs.~\cite{Ginelli,SenPNAS13,2010Ramaswamy}, and secondly, asymmetric rods, 
with $U^{\ast}=3$, to mimic the swim-off effect observed for colliding bacteria. In the following, 
we refer to these two situations as
models with weak or strong alignment. The self-propulsion is introduced 
by an effective self-propulsion force $F_0$ which is directed along the main rod axis leading to a constant propulsion
velocity $v_0$~\cite{BtHComment}. Hereby, we do not resolve any 
details about the actual propulsion mechanism.

Colloidal microswimmers move in the low Reynolds number regime, the corresponding overdamped equations of motion 
for the positions  and orientations are 

\begin{eqnarray}
{\bf f }_{\cal T} \cdot \partial_{t} \br_{\alpha}(t) &=&  -\nabla_{\br_{\alpha}}
 U(t) +  F_{0} \bhu_{\alpha}(t), \\
{\bf f}_{\cal{R}} \cdot \partial_{t} \bhu_{\alpha}(t) &=&
-\nabla_{\bhu_{\alpha}} U(t),
\label{eom}
\end{eqnarray}

in terms of the total potential energy  
$U=(1/2)\sum_{\alpha, \beta (\alpha \neq \beta)} U_{\alpha \beta} + \sum_{\alpha} U_{\alpha <}$ 
with $U_{\alpha <}$ the potential energy of rod $\alpha$ with the carrier (the susbcript $<$ is associated with the carrier). 
The one-body translational and rotational friction tensors for the rods ${\bf f}_{\cal T}$ 
and ${\bf f}_{\cal R}$ can be decomposed into parallel $f_\parallel$, perpendicular $f_\perp$
and rotational $f_{\cal R}$ contributions which depend solely on the aspect ratio $a = \ell/\lambda $~\cite{tirado}

\begin{eqnarray}
\frac{2\pi}{f_{||}} &=& \ln a - 0.207 + 0.980a^{-1} - 0.133a^{-2},
\\
\frac{4\pi}{f_{\perp}}&=&\ln a+0.839 + 0.185a^{-1} + 0.233a^{-2},
\\
\frac{\pi a^2}{3f_{\mathcal{R}}} &=& \ln a - 0.662 + 0.917a^{-1} - 0.050a^{-2}.
\end{eqnarray}

The resulting self-propulsion speed $v_0=F_0 / f_{||} $ sets to characteristic time unit $\tau=\ell/v_0$. We
 ignore any thermal fluctuations.

According to previous experiments~\cite{KaiserSokolov_2014} the motion of the submersed carrier will be restricted to 
translation by using an external magnetic field which keeps the orientation of the carrier fixed.
 The carrier-swimmer interaction is implemented analogously to the swimmers by tiling 
the contour length $L$ into Yukawa segments. Most of our data are obtained for
$L=26\ell$ but we do also vary the contour length $L$.

The resulting equation of motion for the carrier is

\begin{eqnarray}
{\bf f }_{<} \cdot \partial_{t} \br_{<}(t) &=&  -\nabla_{\br_{<}} U_{< \alpha}(t),
\end{eqnarray} 
where ${\bf f }_{<}$ corresponds to the hydrodynamic friction tensor of the wedgelike carrier, calculated for the
specific geometry of the carrier using  the software package \texttt{HYDRO++}~\cite{delaTorreNMDC1994,Carrasco99}.

We use a square simulation box with area $A=(3L/\sqrt{2})^2$ and periodic boundary conditions in both directions.
The total number of swimmers is determined by $N= A \phi / \lambda \ell$, where $\phi$ is a dimensionless
packing fraction.


\section{Bulk behaviour}
\label{Sec:Bulk}

\begin{figure}
\centering
\resizebox{1.0\columnwidth}{!}{\includegraphics{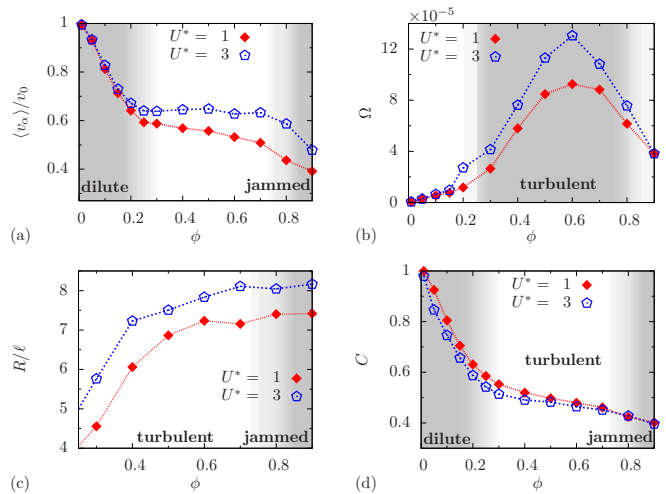}}
\caption{Comparison of different bulk quantities for system with weak (open symbols) and strong (filled symbols) alignment. 
(a) Averaged reduced swimmer velocity $\langle v_\alpha \rangle/ v_0$, (b) mean vorticity $\Omega$,  
(c) typical reduced swirl size  $R/\ell$, and (d) alignment coefficient $C$ as a function of swimmer packing fraction. 
Shaded areas indicate the emergent dynamical states.}
\label{fig:Bulk}   
\end{figure}

Let us start with the characterization of the emergent dynamical states for both considered situations in absence of the carrier. 
As suitable order parameters we use the averaged swimming speed $\langle v_{\alpha} \rangle / v_0$, measured via
 the mean swimmer displacement 
during a time step $\Delta t=10^{-3}\tau$, and the enstrophy  
$\Omega = \frac{1}{2} \langle | [\nabla \times \textbf{V}(\br,t)] \cdot \hat{{\bf e}}_{z}| ^{2} \rangle$ 
for a velocity field $\textbf{V}(\br,t)$ coarse-grained in space over three swimmer lengths.  The results are shown in 
Figs.~\ref{fig:Bulk}(a),(b) and show qualitatively the same behavior for both situations, though the achieved values 
for $\langle v_{\alpha} \rangle / v_0$ and $\Omega$  are slightly higher for swimmers with less alignment. 
In agreement with experiments~\cite{PNAS,2009SoAr}, we can distinguish between three 
dynamical states as a function of increasing swimmer density. 
For low swimmer packing fractions, $\phi \lesssim 0.25$, the average swimmer velocity is 
$\langle v_{\alpha} \rangle \gtrsim 0.6v_0$ due to small amount of collisions, leading to a {\it dilute} state. 
For larger densities $0.25 \lesssim \phi \lesssim 0.75$ the velocity is almost constant and the system reveals a large 
enstrophy $\Omega$ for both systems. Since the enstrophy is a  convenient indicator for bacterial 
turbulence~\cite{PNAS,WensinkJPCM}, we will refer to this state as {\it turbulent}.  For high densities $0.75 \lesssim \phi$ the 
system becomes dynamically {\it jammed}, $\langle v_{\alpha} \rangle \lesssim 0.5v_0$. Using the equal-time spatial 
velocity autocorrelation function, we can determine the typical swirl radius $R$ for various swimmer concentrations 
by its first minimum~\cite{WensinkJPCM}. Hereby, in case of the asymmetric particle model, the typical swirls size 
is larger, see Fig.~\ref{fig:Bulk}(c).

To quantify the influence of the body shape asymmetry on the (collective) motion of the swimmers, we study the parallelism
of the coarse-grained velocity field and the similarly coarse-grained  
orientation field $\textbf{U}(\br,t)$ using the coefficient $C$

\begin{eqnarray}
C = \frac{\langle \cos \theta \rangle - 2/\pi}{1-2/\pi},
\end{eqnarray}
where $\theta$ is the angle between both fields, see Fig.~\ref{fig:Bulk}(d). Perfectly parallel fields lead to $C=1$, while 
random directions reveal $C=0$,  hence $\left< \cos \theta  \right>=2/\pi$ for $\theta \in \{-\pi/2,\pi/2\}$.
In case of the symmetric rods, the mutual collisions in the dilute regime already form dense aligned clusters, leading to 
 high parallelism of the two fields.  With increasing swimmer densities and the emergence of
 large scaled swirls the coefficient $C$ decreases. As is intuitively expected, the stronger the interaction alignment
the larger the parallelism between the two fields.

 
\section{Transport of a wedgelike carrier}
\label{Sec:Transport}
\subsection{Transport efficiency}

\begin{figure}
\centering
\resizebox{1.0\columnwidth}{!}{\includegraphics{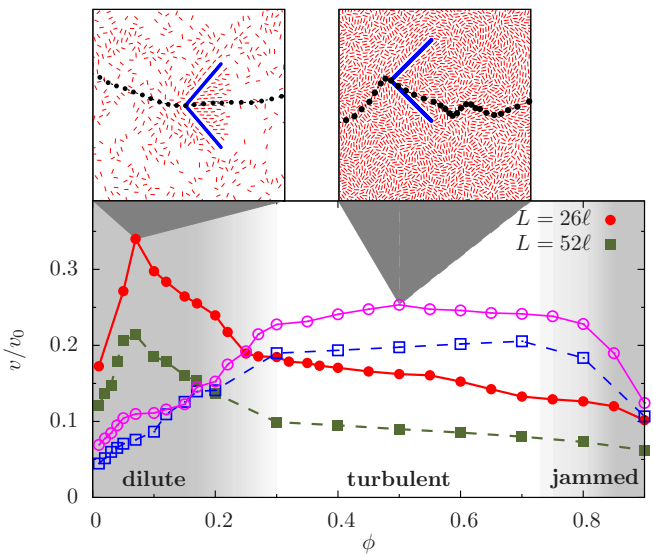}}
\caption{Transport efficiency for various swimmer densities and two contour lengths of the carrier, $L=26 \ell$ (circles) and 
$L=52\ell$ (squares).  Filled symbols correspond to weak and open ones to strong alignment. The 
dynamical states are indicated by shaded areas.  The insets show the temporal progress of the carrier position indicated by the 
dotted line.}
\label{fig:Transport}   
\end{figure}

Now, we study the transport efficiency $v/v_0$ of a wedgelike carrier  for both alignment situations.
Due to the symmetry of the wedge any averaged directed motion perpendicular to the apex will vanish. In the apex direction there 
is no such symmetry and due to rectification the carrier will propagate along this direction. The resulting transport efficiency 
$v/v_0$ is shown in Fig.~\ref{fig:Transport} for two selected contour lengths of the carrier.

While the maximal transport efficiency in around 0.25 for the asymmetric particles, as confirmed by experiment
~\cite{KaiserSokolov_2014}, the efficiency for symmetric swimmer is  larger $v/v_0 \approx 0.35$. Moreover, the characteristic
density for which the transport is optimal is vastly different: it occurs at low densities in the dilute regime for strong
alignment but is significantly shifted towards the turbulent regime for weak alignment.
 However, both aliment conditions reveal an almost constant efficiency in the turbulent regime. Finally, if the active 
fluids jam, the carrier velocity clearly decreases as well.

\subsection{Optimization of wedge shape}
\begin{figure}
\centering
\resizebox{1.0\columnwidth}{!}{\includegraphics{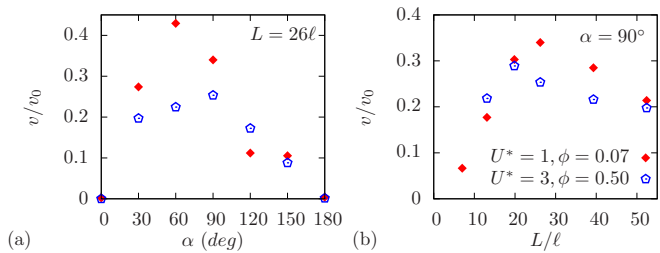}}
\caption{Transport speed for (a) varied apex angle and fixed contour length $L=26\ell$ and (b) varied length and fixed apex angle $\alpha=90^{\circ}$ for both situations at a fixed given swimmer packing fraction $\phi$.}
\label{fig:MaxTransport}   
\end{figure}

We now study the carrier geometry leading to maximal transport efficiency. First of all, we choose for both situations 
the density which showed the highest transport and vary the apex angle for a fixed contour length $L=26\ell$, see Fig.
~\ref{fig:MaxTransport}(a). Clearly, there has to be an optimal angle between zero and 180 degrees 
as these two extreme cases do not lead to any transport at all.
As a result, the optimal wedge for weak alignment has an apex 
angle $\alpha = 90^{\circ}$ while the optimal angle in the strong alignment model is around  $\alpha = 60^{\circ}$.

Complementarily, we then have modified the length for a given 
apex angle $\alpha = 90^{\circ}$, see Fig.~\ref{fig:MaxTransport}(b).
We find an  optimal length  of $L\approx 20 \ell$  in the weak alignment model and a larger length
of $L\approx 26 \ell$ in the strong alignment model.

\subsection{Transport mechanism}

\begin{figure}
\centering
\resizebox{1.0\columnwidth}{!}{\includegraphics{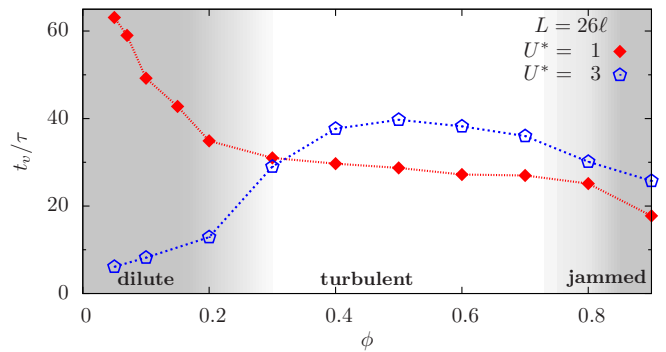}}
\caption{Velocity correlation time $t_v/\tau$ for both particle models, a carrier with contour length $L=26\ell$ and varied swimmer packing fraction.}
\label{fig:CorrTime}   
\end{figure}

First of all, we recapitulate the transport mechanisms for the weak alignment model which has been recently explained 
by swirl shielding inside the carrier~\cite{KaiserSokolov_2014}.
When turbulence sets in, there is a shielding of turbulent fluctuation near the walls of the carrier. Due 
to the wedgelike geometry this shielding is more pronounced inside the carrier than outside and thus
a shielded area near the cusp emerges.
Swimmers are trapped inside this area for a long time and are able to push and thereby 
transport the wedge since they are rectified by the carrier.
This process is limited by the flipping processes of the pushing microswimmers inside the wedge.
The latter gives rise to fluctuations in the carrier velocity which are characterized by a typical
correlation time scale $t_v$. This time $t_v$ is set by the 
decay of the normalized and shifted carrier velocity autocorrelation function defined as
\bea
C_v(t) = \frac{\langle v(t_0) v(t_0 + t) \rangle - \langle v\rangle ^2}{\langle v^2 \rangle - \langle v\rangle ^2}.
\eea
A numerical fit reveals that this quantity decays as $\exp(-t/t_v)$ from which the typical correlation time 
$t_v$ can be extracted.
The results are plotted 
in Fig.~\ref{fig:CorrTime} for the two alignment situations considered in this work.
There is no such swirl shielded area outside the wedge as the swirls can sweep all swimmers away.
This imbalance after all pushes the carrier forward. Clearly, this
 mechanism is valid for both alignment situations in the turbulent state. 
However, the transport efficiency achieved  (see again Fig.~\ref{fig:Transport}) is larger 
for weak alignment since  the typical swirl size is larger than in the strong alignment case (see again Fig.~\ref{fig:Bulk})
 and therefore the corresponding swirl-shielded area is larger leaving more space for pushing microswimmers.
As a consequence of  the swirl shielding concept, the optimal transport  is achieved when the apex width of the 
carrier is comparable to the typical swirl size, $L = 2\sqrt{2}R \approx 21 \ell$, see again Fig.~\ref{fig:MaxTransport}.

\begin{figure}
\centering
\resizebox{1.0\columnwidth}{!}{\includegraphics{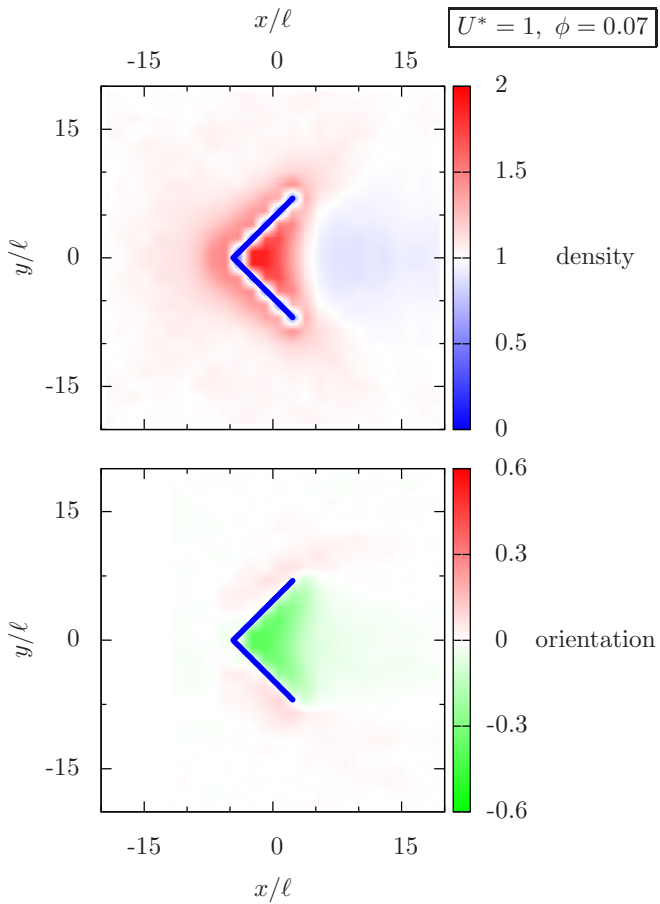}}
\caption{Intensity plots for (top) the local swimmer density around the carrier and (bottom) the averaged swimmer orientations, 
using $\langle \cos \varphi \rangle$  for the situation of strong alignment and a packing fraction $\phi=0.07$.}
\label{fig:IntensityPlots}   
\end{figure}

For strong alignment interactions, the swirl-shielding concept is overwhelmed by another mechanism which 
occurs already in the dilute regime.  Figure~\ref{fig:IntensityPlots} shows  the local density and the
swimmer orientation (by plotting the average $\langle \cos \varphi \rangle$  around the carrier). 
The density exhibits a "hot spot" near 
the cusp and a depleted zone in the wake of the carrier. Due to the directed transport the carrier acts like a {\it bulldozer}
and accumulates swimmers in its front. The intensity plot for the swimmer orientation shows a clear rectification of swimmers 
within the wedge even in the wake, where a huge smectic cluster is formed, see left inset of Fig.~\ref{fig:Transport}.
This cluster is very stable as indicated by a large correlation time $t_v$ in the carrier velocity, see again Fig.~\ref{fig:CorrTime},
leading to persistent straight motion of the carrier.

\begin{figure}
\centering
\resizebox{1.0\columnwidth}{!}{\includegraphics{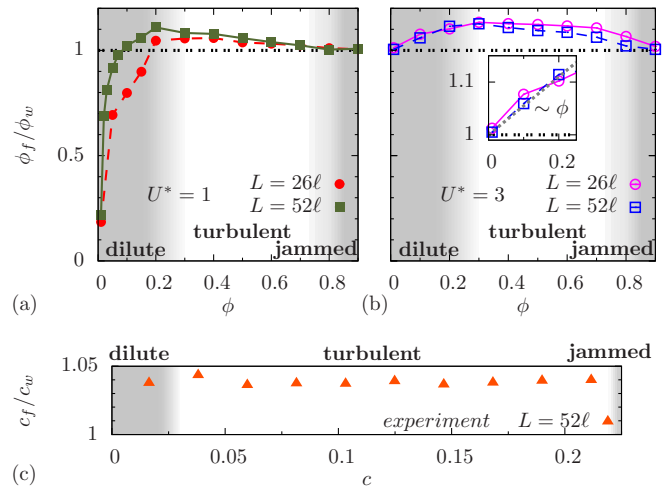}}
\caption{Comparison of the density around the carrier using the ratio $\phi_f/\phi_w$, i.e. the ratio
of the two densities in the front and in the wake for both interaction models, the two selected contour length as before and varied swimmer packing fractions $\phi$. The inset in (b) shows a close up for the dilute state and a linear dependence of the ratio on the packing fraction $\phi$. (c) Experimentally obtained concentration ratio $c_f/c_w$, with $c$ the
three-dimensional volume fraction.}
\label{fig:Density}   
\end{figure}

To supplement this picture, we finally present data for the local average density  in front of the carrier ($\phi_f$) 
and in its wake ($\phi_w$) in Fig.~\ref{fig:Density}.
The emergent smectic cluster in the wake of the carrier for the strong alignment model leads to 
$\phi_f / \phi_w \rightarrow 0$ for small swimmer densities, see Fig.~\ref{fig:Density}(a) 
which supports the strong pushing efficiency. When collective motion starts in the bulk, the ratio $\phi_f / \phi_w$
gets  larger than unity implying that the transport efficiency is decreased.
For the weak alignment model this ratio is always larger than unity, 
see Fig.~\ref{fig:Density}(b). 
This can be confirmed by experiments on {\it Bacillus subtilis} and a microwedge
with a contour length $L=52\ell$~\cite{KaiserSokolov_2014}, see Fig.~\ref{fig:Density}(c).
We can predict this ratio by a simple scaling argument in the dilute regime.
The carrier velocity is $v \sim v_0 \phi$ and the achieved directed motion leads to a density 
gradient $[v \sim D \nabla \phi$, with $D \sim v_0 \ell / \phi_w]$.
According to this, we can approximate the resulting density ratio for
a moving carrier by $\phi_f / \phi_w - 1 \sim \phi$, which is shown in the  inset of Fig.~\ref{fig:Density}(b).

To summarize, there are two different mechanisms at work for optimal carrier transport, namely swirl 
shielding for weak aligning interactions and a large polar pushing cloud of swimmer for strong alignment 
interactions. The first occurs in the turbulent regime while the latter is in the dilute regime.


\section{Conclusion}
\label{conc}
In conclusion, we have demonstrated that the actual transport speed of a passive microwedge ("carrier")
immersed into an active bath depends
on the aligning properties of the individual microswimmers. For strong aligning interactions, as realized 
for artificial rod-like  microswimmers, a polar oriented cloud in the wake of the carrier pushes the carrier forward.
Conversely, for interaction without strong alignment, as realized for bacterial swimmers, the most efficient transport
occurs in the turbulent state of the active bath and is caused by swirl depletion.
Our results were obtained by computer simulations and can be verified in experiments. In particular,
experiment with artificial microswimmers exhibiting a strong aligning interaction are highly desirably
to test the predicted picture of a polar-ordered wake.

Future work should focus on different shapes of carriers like $L$-particles or $C$-particles 
which have been considered theoretically~\cite{MalloryArxiv,Wensink2014} but should be realized in experiments.
Finally, the carrier transport can be possibly used as building block to fabricate more complicated
micro- and nanomachines steered by an active bath.

\acknowledgments
We thank Borge ten Hagen for helpful discussions. A.K. was supported by the ERC Advanced Grant INTERCOCOS (Grant No. 267499)
and H.L. by the SPP 1726 of the DFG. Work by A.S. and I.S.A. was supported by the U.S. Department of Energy (DOE),
Office of Science, Basic Energy Sciences (BES), Materials Science and Engineering Devision. 

\bibliography{refs}

\begin{thebibliography}{99}%
\makeatletter
\providecommand \@ifxundefined [1]{%
 \@ifx{#1\undefined}
}%
\providecommand \@ifnum [1]{%
 \ifnum #1\expandafter \@firstoftwo
 \else \expandafter \@secondoftwo
 \fi
}%
\providecommand \@ifx [1]{%
 \ifx #1\expandafter \@firstoftwo
 \else \expandafter \@secondoftwo
 \fi
}%
\providecommand \natexlab [1]{#1}%
\providecommand \enquote  [1]{``#1''}%
\providecommand \bibnamefont  [1]{#1}%
\providecommand \bibfnamefont [1]{#1}%
\providecommand \citenamefont [1]{#1}%
\providecommand \href@noop [0]{\@secondoftwo}%
\providecommand \href [0]{\begingroup \@sanitize@url \@href}%
\providecommand \@href[1]{\@@startlink{#1}\@@href}%
\providecommand \@@href[1]{\endgroup#1\@@endlink}%
\providecommand \@sanitize@url [0]{\catcode `\\12\catcode `\$12\catcode
  `\&12\catcode `\#12\catcode `\^12\catcode `\_12\catcode `\%12\relax}%
\providecommand \@@startlink[1]{}%
\providecommand \@@endlink[0]{}%
\providecommand \url  [0]{\begingroup\@sanitize@url \@url }%
\providecommand \@url [1]{\endgroup\@href {#1}{\urlprefix }}%
\providecommand \urlprefix  [0]{URL }%
\providecommand \Eprint [0]{\href }%
\providecommand \doibase [0]{http://dx.doi.org/}%
\providecommand \selectlanguage [0]{\@gobble}%
\providecommand \bibinfo  [0]{\@secondoftwo}%
\providecommand \bibfield  [0]{\@secondoftwo}%
\providecommand \translation [1]{[#1]}%
\providecommand \BibitemOpen [0]{}%
\providecommand \bibitemStop [0]{}%
\providecommand \bibitemNoStop [0]{.\EOS\space}%
\providecommand \EOS [0]{\spacefactor3000\relax}%
\providecommand \BibitemShut  [1]{\csname bibitem#1\endcsname}%
\let\auto@bib@innerbib\@empty
\bibitem [{\citenamefont {Romanczuk}\ \emph {et~al.}(2012)\citenamefont
  {Romanczuk}, \citenamefont {B\"ar}, \citenamefont {Ebeling}, \citenamefont
  {Linder},\ and\ \citenamefont {Schimansky-Geier}}]{Romanczuk2012}%
  \BibitemOpen
  \bibfield  {author} {\bibinfo {author} {\bibfnamefont {P.}~\bibnamefont
  {Romanczuk}}, \bibinfo {author} {\bibfnamefont {M.}~\bibnamefont {B\"ar}},
  \bibinfo {author} {\bibfnamefont {W.}~\bibnamefont {Ebeling}}, \bibinfo
  {author} {\bibfnamefont {B.}~\bibnamefont {Linder}}, \ and\ \bibinfo {author}
  {\bibfnamefont {L.}~\bibnamefont {Schimansky-Geier}},\ }\href {\doibase
  10.1140/epjst/e2012-01529-y} {\bibfield  {journal} {\bibinfo  {journal} {Eur.
  Phys. J. Spec. Top.}\ }\textbf {\bibinfo {volume} {202}},\ \bibinfo {pages}
  {1} (\bibinfo {year} {2012})}\BibitemShut {NoStop}%
\bibitem [{\citenamefont {Cates}(2012)}]{Cates_2012}%
  \BibitemOpen
  \bibfield  {author} {\bibinfo {author} {\bibfnamefont {M.~E.}\ \bibnamefont
  {Cates}},\ }\href {\doibase 10.1088/0034-4885/75/4/042601} {\bibfield
  {journal} {\bibinfo  {journal} {Rep. Prog. Phys.}\ }\textbf {\bibinfo
  {volume} {75}},\ \bibinfo {pages} {042601} (\bibinfo {year}
  {2012})}\BibitemShut {NoStop}%
\bibitem [{\citenamefont {Marchetti}\ \emph {et~al.}(2013)\citenamefont
  {Marchetti}, \citenamefont {Joanny}, \citenamefont {Ramaswamy}, \citenamefont
  {Liverpool}, \citenamefont {Prost}, \citenamefont {Rao},\ and\ \citenamefont
  {Simha}}]{Marchetti_Rev}%
  \BibitemOpen
  \bibfield  {author} {\bibinfo {author} {\bibfnamefont {M.~C.}\ \bibnamefont
  {Marchetti}}, \bibinfo {author} {\bibfnamefont {J.~F.}\ \bibnamefont
  {Joanny}}, \bibinfo {author} {\bibfnamefont {S.}~\bibnamefont {Ramaswamy}},
  \bibinfo {author} {\bibfnamefont {T.~B.}\ \bibnamefont {Liverpool}}, \bibinfo
  {author} {\bibfnamefont {J.}~\bibnamefont {Prost}}, \bibinfo {author}
  {\bibfnamefont {M.}~\bibnamefont {Rao}}, \ and\ \bibinfo {author}
  {\bibfnamefont {R.~A.}\ \bibnamefont {Simha}},\ }\href {\doibase
  10.1103/RevModPhys.85.1143} {\bibfield  {journal} {\bibinfo  {journal} {Rev.
  Mod. Phys.}\ }\textbf {\bibinfo {volume} {85}},\ \bibinfo {pages} {1143}
  (\bibinfo {year} {2013})}\BibitemShut {NoStop}%
\bibitem [{\citenamefont {Aranson}(2013{\natexlab{a}})}]{aranson_ufn}%
  \BibitemOpen
  \bibfield  {author} {\bibinfo {author} {\bibfnamefont {I.~S.}\ \bibnamefont
  {Aranson}},\ }\href@noop {} {\bibfield  {journal} {\bibinfo  {journal}
  {Physics-Uspekhi}\ }\textbf {\bibinfo {volume} {56}},\ \bibinfo {pages} {79}
  (\bibinfo {year} {2013}{\natexlab{a}})}\BibitemShut {NoStop}%
\bibitem [{\citenamefont {Sokolov}\ \emph {et~al.}(2007)\citenamefont
  {Sokolov}, \citenamefont {Aranson}, \citenamefont {Kessler},\ and\
  \citenamefont {Goldstein}}]{2007SoEtAl}%
  \BibitemOpen
  \bibfield  {author} {\bibinfo {author} {\bibfnamefont {A.}~\bibnamefont
  {Sokolov}}, \bibinfo {author} {\bibfnamefont {I.~S.}\ \bibnamefont
  {Aranson}}, \bibinfo {author} {\bibfnamefont {J.~O.}\ \bibnamefont
  {Kessler}}, \ and\ \bibinfo {author} {\bibfnamefont {R.~E.}\ \bibnamefont
  {Goldstein}},\ }\href {\doibase 10.1103/PhysRevLett.98.158102} {\bibfield
  {journal} {\bibinfo  {journal} {Phys. Rev. Lett.}\ }\textbf {\bibinfo
  {volume} {98}},\ \bibinfo {pages} {158102} (\bibinfo {year}
  {2007})}\BibitemShut {NoStop}%
\bibitem [{\citenamefont {Shenoy}\ \emph {et~al.}(2007)\citenamefont {Shenoy},
  \citenamefont {Tambe}, \citenamefont {Prasad},\ and\ \citenamefont
  {Theriot}}]{Shenoy_PNAS}%
  \BibitemOpen
  \bibfield  {author} {\bibinfo {author} {\bibfnamefont {V.~B.}\ \bibnamefont
  {Shenoy}}, \bibinfo {author} {\bibfnamefont {D.~T.}\ \bibnamefont {Tambe}},
  \bibinfo {author} {\bibfnamefont {A.}~\bibnamefont {Prasad}}, \ and\ \bibinfo
  {author} {\bibfnamefont {J.~A.}\ \bibnamefont {Theriot}},\ }\href {\doibase
  10.1073/pnas.0702454104} {\bibfield  {journal} {\bibinfo  {journal} {Proc.
  Natl. Acad. Sci. U.S.A.}\ }\textbf {\bibinfo {volume} {104}},\ \bibinfo
  {pages} {8229} (\bibinfo {year} {2007})}\BibitemShut {NoStop}%
\bibitem [{\citenamefont {Schmidt}\ \emph {et~al.}(2008)\citenamefont
  {Schmidt}, \citenamefont {van~der Gucht}, \citenamefont {Biesheuvel},
  \citenamefont {Weinkamer}, \citenamefont {Helfer},\ and\ \citenamefont
  {Frey}}]{Schmidt2008}%
  \BibitemOpen
  \bibfield  {author} {\bibinfo {author} {\bibfnamefont {S.}~\bibnamefont
  {Schmidt}}, \bibinfo {author} {\bibfnamefont {J.}~\bibnamefont {van~der
  Gucht}}, \bibinfo {author} {\bibfnamefont {P.~M.}\ \bibnamefont
  {Biesheuvel}}, \bibinfo {author} {\bibfnamefont {R.}~\bibnamefont
  {Weinkamer}}, \bibinfo {author} {\bibfnamefont {E.}~\bibnamefont {Helfer}}, \
  and\ \bibinfo {author} {\bibfnamefont {A.}~\bibnamefont {Frey}},\ }\href
  {\doibase 10.1007/s00249-008-0340-x} {\bibfield  {journal} {\bibinfo
  {journal} {Europ. Biophys. J.}\ }\textbf {\bibinfo {volume} {37}},\ \bibinfo
  {pages} {1361} (\bibinfo {year} {2008})}\BibitemShut {NoStop}%
\bibitem [{\citenamefont {Schwarz-Linek}\ \emph {et~al.}(2012)\citenamefont
  {Schwarz-Linek}, \citenamefont {Valeriani}, \citenamefont {Cacciuto},
  \citenamefont {Cates}, \citenamefont {Marenduzzo}, \citenamefont {Morozov},\
  and\ \citenamefont {Poon}}]{Poon}%
  \BibitemOpen
  \bibfield  {author} {\bibinfo {author} {\bibfnamefont {J.}~\bibnamefont
  {Schwarz-Linek}}, \bibinfo {author} {\bibfnamefont {C.}~\bibnamefont
  {Valeriani}}, \bibinfo {author} {\bibfnamefont {A.}~\bibnamefont {Cacciuto}},
  \bibinfo {author} {\bibfnamefont {M.~E.}\ \bibnamefont {Cates}}, \bibinfo
  {author} {\bibfnamefont {D.}~\bibnamefont {Marenduzzo}}, \bibinfo {author}
  {\bibfnamefont {A.~N.}\ \bibnamefont {Morozov}}, \ and\ \bibinfo {author}
  {\bibfnamefont {W.~C.~K.}\ \bibnamefont {Poon}},\ }\href {\doibase
  10.1073/pnas.1116334109} {\bibfield  {journal} {\bibinfo  {journal} {Proc.
  Natl. Acad. Sci. U.S.A.}\ }\textbf {\bibinfo {volume} {109}},\ \bibinfo
  {pages} {4052} (\bibinfo {year} {2012})}\BibitemShut {NoStop}%
\bibitem [{\citenamefont {Polin}\ \emph {et~al.}(2009)\citenamefont {Polin},
  \citenamefont {Tuval}, \citenamefont {Drescher}, \citenamefont {Gollub},\
  and\ \citenamefont {Goldstein}}]{SwimmingAlga}%
  \BibitemOpen
  \bibfield  {author} {\bibinfo {author} {\bibfnamefont {M.}~\bibnamefont
  {Polin}}, \bibinfo {author} {\bibfnamefont {I.}~\bibnamefont {Tuval}},
  \bibinfo {author} {\bibfnamefont {K.}~\bibnamefont {Drescher}}, \bibinfo
  {author} {\bibfnamefont {J.~P.}\ \bibnamefont {Gollub}}, \ and\ \bibinfo
  {author} {\bibfnamefont {R.~E.}\ \bibnamefont {Goldstein}},\ }\href {\doibase
  10.1126/science.1172667} {\bibfield  {journal} {\bibinfo  {journal}
  {Science}\ }\textbf {\bibinfo {volume} {325}},\ \bibinfo {pages} {487}
  (\bibinfo {year} {2009})}\BibitemShut {NoStop}%
\bibitem [{\citenamefont {Guasto}\ \emph {et~al.}(2010)\citenamefont {Guasto},
  \citenamefont {Johnson},\ and\ \citenamefont {Gollub}}]{CRein2010}%
  \BibitemOpen
  \bibfield  {author} {\bibinfo {author} {\bibfnamefont {J.~S.}\ \bibnamefont
  {Guasto}}, \bibinfo {author} {\bibfnamefont {K.~A.}\ \bibnamefont {Johnson}},
  \ and\ \bibinfo {author} {\bibfnamefont {J.~P.}\ \bibnamefont {Gollub}},\
  }\href@noop {} {\bibfield  {journal} {\bibinfo  {journal} {Phys. Rev. Lett.}\
  }\textbf {\bibinfo {volume} {105}},\ \bibinfo {pages} {168102} (\bibinfo
  {year} {2010})}\BibitemShut {NoStop}%
\bibitem [{\citenamefont {Ma}\ \emph {et~al.}(2014)\citenamefont {Ma},
  \citenamefont {Klindt}, \citenamefont {Riedel-Kruse}, \citenamefont
  {J\"ulicher},\ and\ \citenamefont {Friedrich}}]{BMF_PRL14}%
  \BibitemOpen
  \bibfield  {author} {\bibinfo {author} {\bibfnamefont {R.}~\bibnamefont
  {Ma}}, \bibinfo {author} {\bibfnamefont {G.~S.}\ \bibnamefont {Klindt}},
  \bibinfo {author} {\bibfnamefont {I.~H.}\ \bibnamefont {Riedel-Kruse}},
  \bibinfo {author} {\bibfnamefont {F.}~\bibnamefont {J\"ulicher}}, \ and\
  \bibinfo {author} {\bibfnamefont {B.~M.}\ \bibnamefont {Friedrich}},\
  }\href@noop {} {\bibfield  {journal} {\bibinfo  {journal} {Phys. Rev. Lett.}\
  }\textbf {\bibinfo {volume} {113}},\ \bibinfo {pages} {048101} (\bibinfo
  {year} {2014})}\BibitemShut {NoStop}%
\bibitem [{\citenamefont {Riedel}\ \emph {et~al.}(2005)\citenamefont {Riedel},
  \citenamefont {Kruse},\ and\ \citenamefont {Howard}}]{2005Riedel_Science}%
  \BibitemOpen
  \bibfield  {author} {\bibinfo {author} {\bibfnamefont {I.~H.}\ \bibnamefont
  {Riedel}}, \bibinfo {author} {\bibfnamefont {K.}~\bibnamefont {Kruse}}, \
  and\ \bibinfo {author} {\bibfnamefont {J.}~\bibnamefont {Howard}},\ }\href
  {\doibase 10.1126/science.1110329} {\bibfield  {journal} {\bibinfo  {journal}
  {Science}\ }\textbf {\bibinfo {volume} {309}},\ \bibinfo {pages} {300}
  (\bibinfo {year} {2005})}\BibitemShut {NoStop}%
\bibitem [{\citenamefont {Friedrich}\ and\ \citenamefont
  {J\"ulicher}(2008)}]{Friedrich}%
  \BibitemOpen
  \bibfield  {author} {\bibinfo {author} {\bibfnamefont {B.~M.}\ \bibnamefont
  {Friedrich}}\ and\ \bibinfo {author} {\bibfnamefont {F.}~\bibnamefont
  {J\"ulicher}},\ }\href {\doibase 10.1088/1367-2630/10/12/123025} {\bibfield
  {journal} {\bibinfo  {journal} {New J. Phys.}\ }\textbf {\bibinfo {volume}
  {10}},\ \bibinfo {pages} {123035} (\bibinfo {year} {2008})}\BibitemShut
  {NoStop}%
\bibitem [{\citenamefont {Woolley}(2003)}]{Woolley}%
  \BibitemOpen
  \bibfield  {author} {\bibinfo {author} {\bibfnamefont {D.~M.}\ \bibnamefont
  {Woolley}},\ }\href {\doibase 10.1530/rep.0.1260259} {\bibfield  {journal}
  {\bibinfo  {journal} {Reproduction}\ }\textbf {\bibinfo {volume} {216}},\
  \bibinfo {pages} {259} (\bibinfo {year} {2003})}\BibitemShut {NoStop}%
\bibitem [{\citenamefont {Vicsek}\ and\ \citenamefont
  {Zafeiris}(2012)}]{Vicsek_Report2012}%
  \BibitemOpen
  \bibfield  {author} {\bibinfo {author} {\bibfnamefont {T.}~\bibnamefont
  {Vicsek}}\ and\ \bibinfo {author} {\bibfnamefont {A.}~\bibnamefont
  {Zafeiris}},\ }\href {\doibase 10.1016/j.physrep.2012.03.004} {\bibfield
  {journal} {\bibinfo  {journal} {Phys. Rep.}\ }\textbf {\bibinfo {volume}
  {517}},\ \bibinfo {pages} {71} (\bibinfo {year} {2012})}\BibitemShut
  {NoStop}%
\bibitem [{\citenamefont {Ballerini}\ \emph {et~al.}(2008)\citenamefont
  {Ballerini}, \citenamefont {Cabibbo}, \citenamefont {Candelier},
  \citenamefont {Cavagna}, \citenamefont {Cisbani}, \citenamefont {Giardina},
  \citenamefont {Lecomte}, \citenamefont {Orlandi}, \citenamefont {Parisi},
  \citenamefont {Procaccini}, \citenamefont {Viale},\ and\ \citenamefont
  {Zdravkovic}}]{BirdsPNAS}%
  \BibitemOpen
  \bibfield  {author} {\bibinfo {author} {\bibfnamefont {M.}~\bibnamefont
  {Ballerini}}, \bibinfo {author} {\bibfnamefont {N.}~\bibnamefont {Cabibbo}},
  \bibinfo {author} {\bibfnamefont {R.}~\bibnamefont {Candelier}}, \bibinfo
  {author} {\bibfnamefont {A.}~\bibnamefont {Cavagna}}, \bibinfo {author}
  {\bibfnamefont {E.}~\bibnamefont {Cisbani}}, \bibinfo {author} {\bibfnamefont
  {I.}~\bibnamefont {Giardina}}, \bibinfo {author} {\bibfnamefont
  {V.}~\bibnamefont {Lecomte}}, \bibinfo {author} {\bibfnamefont
  {A.}~\bibnamefont {Orlandi}}, \bibinfo {author} {\bibfnamefont
  {G.}~\bibnamefont {Parisi}}, \bibinfo {author} {\bibfnamefont
  {A.}~\bibnamefont {Procaccini}}, \bibinfo {author} {\bibfnamefont
  {M.}~\bibnamefont {Viale}}, \ and\ \bibinfo {author} {\bibfnamefont
  {V.}~\bibnamefont {Zdravkovic}},\ }\href {\doibase 10.1073/pnas.0711437105}
  {\bibfield  {journal} {\bibinfo  {journal} {Proc. Natl. Acad. Sci. U.S.A.}\
  }\textbf {\bibinfo {volume} {105}},\ \bibinfo {pages} {1232} (\bibinfo {year}
  {2008})}\BibitemShut {NoStop}%
\bibitem [{\citenamefont {Katz}\ \emph {et~al.}(2011)\citenamefont {Katz},
  \citenamefont {Tunstr{\o}m}, \citenamefont {Ioannou}, \citenamefont {Huepe},\
  and\ \citenamefont {Couzin}}]{FishPNAS}%
  \BibitemOpen
  \bibfield  {author} {\bibinfo {author} {\bibfnamefont {Y.}~\bibnamefont
  {Katz}}, \bibinfo {author} {\bibfnamefont {K.}~\bibnamefont {Tunstr{\o}m}},
  \bibinfo {author} {\bibfnamefont {C.~C.}\ \bibnamefont {Ioannou}}, \bibinfo
  {author} {\bibfnamefont {C.}~\bibnamefont {Huepe}}, \ and\ \bibinfo {author}
  {\bibfnamefont {I.~D.}\ \bibnamefont {Couzin}},\ }\href {\doibase
  10.1073/pnas.1107583108} {\bibfield  {journal} {\bibinfo  {journal} {Proc.
  Natl. Acad. Sci. U.S.A.}\ }\textbf {\bibinfo {volume} {108}},\ \bibinfo
  {pages} {18720} (\bibinfo {year} {2011})}\BibitemShut {NoStop}%
\bibitem [{\citenamefont {Buhl}\ \emph {et~al.}(2006)\citenamefont {Buhl},
  \citenamefont {Sumpter}, \citenamefont {Couzin}, \citenamefont {Hale},
  \citenamefont {Despland}, \citenamefont {Miller},\ and\ \citenamefont
  {Simpson}}]{locusts}%
  \BibitemOpen
  \bibfield  {author} {\bibinfo {author} {\bibfnamefont {J.}~\bibnamefont
  {Buhl}}, \bibinfo {author} {\bibfnamefont {D.~J.~T.}\ \bibnamefont
  {Sumpter}}, \bibinfo {author} {\bibfnamefont {I.~D.}\ \bibnamefont {Couzin}},
  \bibinfo {author} {\bibfnamefont {J.~J.}\ \bibnamefont {Hale}}, \bibinfo
  {author} {\bibfnamefont {E.}~\bibnamefont {Despland}}, \bibinfo {author}
  {\bibfnamefont {E.~R.}\ \bibnamefont {Miller}}, \ and\ \bibinfo {author}
  {\bibfnamefont {S.~J.}\ \bibnamefont {Simpson}},\ }\href {\doibase
  10.1126/science.1125142} {\bibfield  {journal} {\bibinfo  {journal}
  {Science}\ }\textbf {\bibinfo {volume} {312}},\ \bibinfo {pages} {1402}
  (\bibinfo {year} {2006})}\BibitemShut {NoStop}%
\bibitem [{\citenamefont {John}\ \emph {et~al.}(2009)\citenamefont {John},
  \citenamefont {Schadschneider}, \citenamefont {Chowdhury},\ and\
  \citenamefont {Nishinari}}]{Ants}%
  \BibitemOpen
  \bibfield  {author} {\bibinfo {author} {\bibfnamefont {A.}~\bibnamefont
  {John}}, \bibinfo {author} {\bibfnamefont {A.}~\bibnamefont
  {Schadschneider}}, \bibinfo {author} {\bibfnamefont {D.}~\bibnamefont
  {Chowdhury}}, \ and\ \bibinfo {author} {\bibfnamefont {K.}~\bibnamefont
  {Nishinari}},\ }\href@noop {} {\bibfield  {journal} {\bibinfo  {journal}
  {Phys. Rev. Lett.}\ }\textbf {\bibinfo {volume} {102}},\ \bibinfo {pages}
  {108001} (\bibinfo {year} {2009})}\BibitemShut {NoStop}%
\bibitem [{\citenamefont {Zhang}\ \emph {et~al.}(2012)\citenamefont {Zhang},
  \citenamefont {Klingsch}, \citenamefont {Schadschneider},\ and\ \citenamefont
  {Seyfried}}]{Schadschneider}%
  \BibitemOpen
  \bibfield  {author} {\bibinfo {author} {\bibfnamefont {J.}~\bibnamefont
  {Zhang}}, \bibinfo {author} {\bibfnamefont {W.}~\bibnamefont {Klingsch}},
  \bibinfo {author} {\bibfnamefont {A.}~\bibnamefont {Schadschneider}}, \ and\
  \bibinfo {author} {\bibfnamefont {A.}~\bibnamefont {Seyfried}},\ }\href
  {\doibase 10.1088/1742-5468/2012/02/P02002} {\bibfield  {journal} {\bibinfo
  {journal} {J. Stat. Mech.}\ }\textbf {\bibinfo {volume} {2012}},\ \bibinfo
  {pages} {P02002} (\bibinfo {year} {2012})}\BibitemShut {NoStop}%
\bibitem [{\citenamefont {Helbing}\ \emph {et~al.}(2000)\citenamefont
  {Helbing}, \citenamefont {Farkas},\ and\ \citenamefont {Vicsek}}]{Helbing}%
  \BibitemOpen
  \bibfield  {author} {\bibinfo {author} {\bibfnamefont {D.}~\bibnamefont
  {Helbing}}, \bibinfo {author} {\bibfnamefont {I.}~\bibnamefont {Farkas}}, \
  and\ \bibinfo {author} {\bibfnamefont {T.}~\bibnamefont {Vicsek}},\ }\href
  {\doibase 10.1038/35035023} {\bibfield  {journal} {\bibinfo  {journal}
  {Nature}\ }\textbf {\bibinfo {volume} {407}},\ \bibinfo {pages} {487}
  (\bibinfo {year} {2000})}\BibitemShut {NoStop}%
\bibitem [{\citenamefont {Silverberg}\ \emph {et~al.}(2013)\citenamefont
  {Silverberg}, \citenamefont {Bierbaum}, \citenamefont {Sethna},\ and\
  \citenamefont {Cohen}}]{Silverberg:13}%
  \BibitemOpen
  \bibfield  {author} {\bibinfo {author} {\bibfnamefont {J.~L.}\ \bibnamefont
  {Silverberg}}, \bibinfo {author} {\bibfnamefont {M.}~\bibnamefont
  {Bierbaum}}, \bibinfo {author} {\bibfnamefont {J.~P.}\ \bibnamefont
  {Sethna}}, \ and\ \bibinfo {author} {\bibfnamefont {I.}~\bibnamefont
  {Cohen}},\ }\href {\doibase 10.1103/PhysRevLett.110.228701} {\bibfield
  {journal} {\bibinfo  {journal} {Phys. Rev. Lett.}\ }\textbf {\bibinfo
  {volume} {110}},\ \bibinfo {pages} {228701} (\bibinfo {year}
  {2013})}\BibitemShut {NoStop}%
\bibitem [{\citenamefont {Ghosh}\ and\ \citenamefont
  {Fischer}(2009)}]{ScrewSwimmer}%
  \BibitemOpen
  \bibfield  {author} {\bibinfo {author} {\bibfnamefont {A.}~\bibnamefont
  {Ghosh}}\ and\ \bibinfo {author} {\bibfnamefont {P.}~\bibnamefont
  {Fischer}},\ }\href {\doibase 10.1021/nl900186w} {\bibfield  {journal}
  {\bibinfo  {journal} {Nano Lett.}\ }\textbf {\bibinfo {volume} {9}},\
  \bibinfo {pages} {2243} (\bibinfo {year} {2009})}\BibitemShut {NoStop}%
\bibitem [{\citenamefont {Dreyfus}\ \emph {et~al.}(2005)\citenamefont
  {Dreyfus}, \citenamefont {Baudry}, \citenamefont {Roper}, \citenamefont
  {Fermigier}, \citenamefont {Stone},\ and\ \citenamefont {Bibette}}]{Bibette}%
  \BibitemOpen
  \bibfield  {author} {\bibinfo {author} {\bibfnamefont {R.}~\bibnamefont
  {Dreyfus}}, \bibinfo {author} {\bibfnamefont {J.}~\bibnamefont {Baudry}},
  \bibinfo {author} {\bibfnamefont {M.~L.}\ \bibnamefont {Roper}}, \bibinfo
  {author} {\bibfnamefont {M.}~\bibnamefont {Fermigier}}, \bibinfo {author}
  {\bibfnamefont {H.~A.}\ \bibnamefont {Stone}}, \ and\ \bibinfo {author}
  {\bibfnamefont {J.}~\bibnamefont {Bibette}},\ }\href {\doibase
  10.1038/nature04090} {\bibfield  {journal} {\bibinfo  {journal} {Nature}\
  }\textbf {\bibinfo {volume} {437}},\ \bibinfo {pages} {862} (\bibinfo {year}
  {2005})}\BibitemShut {NoStop}%
\bibitem [{\citenamefont {Tierno}\ \emph {et~al.}(2008)\citenamefont {Tierno},
  \citenamefont {Golestanian}, \citenamefont {Pagonabarraga},\ and\
  \citenamefont {F.~Sagues}}]{Ignacio}%
  \BibitemOpen
  \bibfield  {author} {\bibinfo {author} {\bibfnamefont {P.}~\bibnamefont
  {Tierno}}, \bibinfo {author} {\bibfnamefont {R.}~\bibnamefont {Golestanian}},
  \bibinfo {author} {\bibfnamefont {I.}~\bibnamefont {Pagonabarraga}}, \ and\
  \bibinfo {author} {\bibfnamefont {F.}~\bibnamefont {F.~Sagues}},\ }\href@noop
  {} {\bibfield  {journal} {\bibinfo  {journal} {J. Phys. Chem. B}\ }\textbf
  {\bibinfo {volume} {112}},\ \bibinfo {pages} {16525} (\bibinfo {year}
  {2008})}\BibitemShut {NoStop}%
\bibitem [{\citenamefont {Khalil}\ \emph {et~al.}(2014)\citenamefont {Khalil},
  \citenamefont {Dijkslag}, \citenamefont {Abelmann},\ and\ \citenamefont
  {Misra}}]{MagnetoSperm}%
  \BibitemOpen
  \bibfield  {author} {\bibinfo {author} {\bibfnamefont {I.~S.~M.}\
  \bibnamefont {Khalil}}, \bibinfo {author} {\bibfnamefont {H.~C.}\
  \bibnamefont {Dijkslag}}, \bibinfo {author} {\bibfnamefont {L.}~\bibnamefont
  {Abelmann}}, \ and\ \bibinfo {author} {\bibfnamefont {S.}~\bibnamefont
  {Misra}},\ }\href {\doibase http://dx.doi.org/10.1063/1.4880035} {\bibfield
  {journal} {\bibinfo  {journal} {Appl. Phys. Lett.}\ }\textbf {\bibinfo
  {volume} {104}},\ \bibinfo {eid} {223701} (\bibinfo {year}
  {2014})}\BibitemShut {NoStop}%
\bibitem [{\citenamefont {Paxton}\ \emph {et~al.}(2004)\citenamefont {Paxton},
  \citenamefont {Kistler}, \citenamefont {Olmeda}, \citenamefont {Sen},
  \citenamefont {St.~Angelo}, \citenamefont {Cao}, \citenamefont {Mallouk},
  \citenamefont {Lammert},\ and\ \citenamefont {Crespi}}]{PaxtonJACS2004}%
  \BibitemOpen
  \bibfield  {author} {\bibinfo {author} {\bibfnamefont {W.~F.}\ \bibnamefont
  {Paxton}}, \bibinfo {author} {\bibfnamefont {K.~C.}\ \bibnamefont {Kistler}},
  \bibinfo {author} {\bibfnamefont {C.~C.}\ \bibnamefont {Olmeda}}, \bibinfo
  {author} {\bibfnamefont {A.}~\bibnamefont {Sen}}, \bibinfo {author}
  {\bibfnamefont {S.~K.}\ \bibnamefont {St.~Angelo}}, \bibinfo {author}
  {\bibfnamefont {Y.}~\bibnamefont {Cao}}, \bibinfo {author} {\bibfnamefont
  {T.~E.}\ \bibnamefont {Mallouk}}, \bibinfo {author} {\bibfnamefont {P.~E.}\
  \bibnamefont {Lammert}}, \ and\ \bibinfo {author} {\bibfnamefont {V.~H.}\
  \bibnamefont {Crespi}},\ }\href {\doibase 10.1021/ja047697z} {\bibfield
  {journal} {\bibinfo  {journal} {J. Am. Chem. Soc.}\ }\textbf {\bibinfo
  {volume} {126}},\ \bibinfo {pages} {13424} (\bibinfo {year}
  {2004})}\BibitemShut {NoStop}%
\bibitem [{\citenamefont {Fournier-Bidoz}\ \emph {et~al.}(2005)\citenamefont
  {Fournier-Bidoz}, \citenamefont {Arsenault}, \citenamefont {Manners},\ and\
  \citenamefont {Ozin}}]{SyntheticRod2005}%
  \BibitemOpen
  \bibfield  {author} {\bibinfo {author} {\bibfnamefont {S.}~\bibnamefont
  {Fournier-Bidoz}}, \bibinfo {author} {\bibfnamefont {A.~C.}\ \bibnamefont
  {Arsenault}}, \bibinfo {author} {\bibfnamefont {I.}~\bibnamefont {Manners}},
  \ and\ \bibinfo {author} {\bibfnamefont {G.~A.}\ \bibnamefont {Ozin}},\
  }\href {\doibase 10.1039/B414896G} {\bibfield  {journal} {\bibinfo  {journal}
  {Chem. Commun.}\ ,\ \bibinfo {pages} {441}} (\bibinfo {year}
  {2005})}\BibitemShut {NoStop}%
\bibitem [{\citenamefont {Howse}\ \emph {et~al.}(2007)\citenamefont {Howse},
  \citenamefont {Jones}, \citenamefont {Ryan}, \citenamefont {Gough},
  \citenamefont {Vafabakhsh},\ and\ \citenamefont {Golestanian}}]{Howse_2007}%
  \BibitemOpen
  \bibfield  {author} {\bibinfo {author} {\bibfnamefont {J.~R.}\ \bibnamefont
  {Howse}}, \bibinfo {author} {\bibfnamefont {R.~A.~L.}\ \bibnamefont {Jones}},
  \bibinfo {author} {\bibfnamefont {A.~J.}\ \bibnamefont {Ryan}}, \bibinfo
  {author} {\bibfnamefont {T.}~\bibnamefont {Gough}}, \bibinfo {author}
  {\bibfnamefont {R.}~\bibnamefont {Vafabakhsh}}, \ and\ \bibinfo {author}
  {\bibfnamefont {R.}~\bibnamefont {Golestanian}},\ }\href {\doibase
  10.1103/PhysRevLett.99.048102} {\bibfield  {journal} {\bibinfo  {journal}
  {Phys. Rev. Lett.}\ }\textbf {\bibinfo {volume} {99}},\ \bibinfo {pages}
  {048102} (\bibinfo {year} {2007})}\BibitemShut {NoStop}%
\bibitem [{\citenamefont {Palacci}\ \emph {et~al.}(2010)\citenamefont
  {Palacci}, \citenamefont {Cottin-Bizonne}, \citenamefont {Ybert},\ and\
  \citenamefont {Bocquet}}]{BocquetPRL2010}%
  \BibitemOpen
  \bibfield  {author} {\bibinfo {author} {\bibfnamefont {J.}~\bibnamefont
  {Palacci}}, \bibinfo {author} {\bibfnamefont {C.}~\bibnamefont
  {Cottin-Bizonne}}, \bibinfo {author} {\bibfnamefont {C.}~\bibnamefont
  {Ybert}}, \ and\ \bibinfo {author} {\bibfnamefont {L.}~\bibnamefont
  {Bocquet}},\ }\href {\doibase 10.1103/PhysRevLett.105.088304} {\bibfield
  {journal} {\bibinfo  {journal} {Phys. Rev. Lett.}\ }\textbf {\bibinfo
  {volume} {105}},\ \bibinfo {pages} {088304} (\bibinfo {year}
  {2010})}\BibitemShut {NoStop}%
\bibitem [{\citenamefont {Kapral}(2013)}]{KapralJCP2013}%
  \BibitemOpen
  \bibfield  {author} {\bibinfo {author} {\bibfnamefont {R.}~\bibnamefont
  {Kapral}},\ }\href {\doibase http://dx.doi.org/10.1063/1.4773981} {\bibfield
  {journal} {\bibinfo  {journal} {J. Chem. Phys.}\ }\textbf {\bibinfo {volume}
  {138}},\ \bibinfo {eid} {020901} (\bibinfo {year} {2013})}\BibitemShut
  {NoStop}%
\bibitem [{\citenamefont {Sanchez}\ \emph {et~al.}(2011)\citenamefont
  {Sanchez}, \citenamefont {Solovev}, \citenamefont {Harazim},\ and\
  \citenamefont {Schmidt}}]{SanchezJACS2011}%
  \BibitemOpen
  \bibfield  {author} {\bibinfo {author} {\bibfnamefont {S.}~\bibnamefont
  {Sanchez}}, \bibinfo {author} {\bibfnamefont {A.~A.}\ \bibnamefont
  {Solovev}}, \bibinfo {author} {\bibfnamefont {S.~M.}\ \bibnamefont
  {Harazim}}, \ and\ \bibinfo {author} {\bibfnamefont {O.~G.}\ \bibnamefont
  {Schmidt}},\ }\href {\doibase 10.1021/ja109627w} {\bibfield  {journal}
  {\bibinfo  {journal} {J. Am. Chem. Soc.}\ }\textbf {\bibinfo {volume}
  {133}},\ \bibinfo {pages} {701} (\bibinfo {year} {2011})}\BibitemShut
  {NoStop}%
\bibitem [{\citenamefont {Jiang}\ \emph {et~al.}(2010)\citenamefont {Jiang},
  \citenamefont {Yoshinaga},\ and\ \citenamefont {Sano}}]{Sano_PRL2010}%
  \BibitemOpen
  \bibfield  {author} {\bibinfo {author} {\bibfnamefont {H.-R.}\ \bibnamefont
  {Jiang}}, \bibinfo {author} {\bibfnamefont {N.}~\bibnamefont {Yoshinaga}}, \
  and\ \bibinfo {author} {\bibfnamefont {M.}~\bibnamefont {Sano}},\ }\href
  {\doibase 10.1103/PhysRevLett.105.268302} {\bibfield  {journal} {\bibinfo
  {journal} {Phys. Rev. Lett.}\ }\textbf {\bibinfo {volume} {105}},\ \bibinfo
  {pages} {268302} (\bibinfo {year} {2010})}\BibitemShut {NoStop}%
\bibitem [{\citenamefont {Volpe}\ \emph {et~al.}(2011)\citenamefont {Volpe},
  \citenamefont {Buttinoni}, \citenamefont {Vogt}, \citenamefont {K\"ummerer},\
  and\ \citenamefont {Bechinger}}]{Bechinger_SM11}%
  \BibitemOpen
  \bibfield  {author} {\bibinfo {author} {\bibfnamefont {G.}~\bibnamefont
  {Volpe}}, \bibinfo {author} {\bibfnamefont {I.}~\bibnamefont {Buttinoni}},
  \bibinfo {author} {\bibfnamefont {D.}~\bibnamefont {Vogt}}, \bibinfo {author}
  {\bibfnamefont {H.-J.}\ \bibnamefont {K\"ummerer}}, \ and\ \bibinfo {author}
  {\bibfnamefont {C.}~\bibnamefont {Bechinger}},\ }\href {\doibase
  10.1039/C1SM05960B} {\bibfield  {journal} {\bibinfo  {journal} {Soft Matter}\
  }\textbf {\bibinfo {volume} {7}},\ \bibinfo {pages} {8810} (\bibinfo {year}
  {2011})}\BibitemShut {NoStop}%
\bibitem [{\citenamefont {K\"ummel}\ \emph {et~al.}(2013)\citenamefont
  {K\"ummel}, \citenamefont {ten Hagen}, \citenamefont {Wittkowski},
  \citenamefont {Buttinoni}, \citenamefont {Eichhorn}, \citenamefont {Volpe},
  \citenamefont {L\"owen},\ and\ \citenamefont {Bechinger}}]{BtH_L_part}%
  \BibitemOpen
  \bibfield  {author} {\bibinfo {author} {\bibfnamefont {F.}~\bibnamefont
  {K\"ummel}}, \bibinfo {author} {\bibfnamefont {B.}~\bibnamefont {ten Hagen}},
  \bibinfo {author} {\bibfnamefont {R.}~\bibnamefont {Wittkowski}}, \bibinfo
  {author} {\bibfnamefont {I.}~\bibnamefont {Buttinoni}}, \bibinfo {author}
  {\bibfnamefont {R.}~\bibnamefont {Eichhorn}}, \bibinfo {author}
  {\bibfnamefont {G.}~\bibnamefont {Volpe}}, \bibinfo {author} {\bibfnamefont
  {H.}~\bibnamefont {L\"owen}}, \ and\ \bibinfo {author} {\bibfnamefont
  {C.}~\bibnamefont {Bechinger}},\ }\href {\doibase
  10.1103/PhysRevLett.110.198302} {\bibfield  {journal} {\bibinfo  {journal}
  {Phys. Rev. Lett.}\ }\textbf {\bibinfo {volume} {110}},\ \bibinfo {pages}
  {198302} (\bibinfo {year} {2013})}\BibitemShut {NoStop}%
\bibitem [{\citenamefont {Thutupalli}\ \emph {et~al.}(2011)\citenamefont
  {Thutupalli}, \citenamefont {Seemann},\ and\ \citenamefont
  {Herminghaus}}]{2011Herminghaus}%
  \BibitemOpen
  \bibfield  {author} {\bibinfo {author} {\bibfnamefont {S.}~\bibnamefont
  {Thutupalli}}, \bibinfo {author} {\bibfnamefont {R.}~\bibnamefont {Seemann}},
  \ and\ \bibinfo {author} {\bibfnamefont {S.}~\bibnamefont {Herminghaus}},\
  }\href {\doibase 10.1088/1367-2630/13/7/073021} {\bibfield  {journal}
  {\bibinfo  {journal} {New J. Phys.}\ }\textbf {\bibinfo {volume} {13}},\
  \bibinfo {pages} {073021} (\bibinfo {year} {2011})}\BibitemShut {NoStop}%
\bibitem [{\citenamefont {Yang}\ \emph {et~al.}(2010)\citenamefont {Yang},
  \citenamefont {Marceau},\ and\ \citenamefont {Gompper}}]{yang-gompper}%
  \BibitemOpen
  \bibfield  {author} {\bibinfo {author} {\bibfnamefont {Y.}~\bibnamefont
  {Yang}}, \bibinfo {author} {\bibfnamefont {V.}~\bibnamefont {Marceau}}, \
  and\ \bibinfo {author} {\bibfnamefont {G.}~\bibnamefont {Gompper}},\ }\href
  {\doibase 10.1103/PhysRevE.82.031904} {\bibfield  {journal} {\bibinfo
  {journal} {Phys. Rev. E}\ }\textbf {\bibinfo {volume} {82}},\ \bibinfo
  {pages} {031904} (\bibinfo {year} {2010})}\BibitemShut {NoStop}%
\bibitem [{\citenamefont {Ginelli}\ \emph {et~al.}(2010)\citenamefont
  {Ginelli}, \citenamefont {Peruani}, \citenamefont {B\"ar},\ and\
  \citenamefont {Chat\'e}}]{Ginelli}%
  \BibitemOpen
  \bibfield  {author} {\bibinfo {author} {\bibfnamefont {F.}~\bibnamefont
  {Ginelli}}, \bibinfo {author} {\bibfnamefont {F.}~\bibnamefont {Peruani}},
  \bibinfo {author} {\bibfnamefont {M.}~\bibnamefont {B\"ar}}, \ and\ \bibinfo
  {author} {\bibfnamefont {H.}~\bibnamefont {Chat\'e}},\ }\href {\doibase
  10.1103/PhysRevLett.104.184502} {\bibfield  {journal} {\bibinfo  {journal}
  {Phys. Rev. Lett.}\ }\textbf {\bibinfo {volume} {104}},\ \bibinfo {pages}
  {184502} (\bibinfo {year} {2010})}\BibitemShut {NoStop}%
\bibitem [{\citenamefont {Chen}\ \emph {et~al.}(2012)\citenamefont {Chen},
  \citenamefont {Dong}, \citenamefont {Be'er}, \citenamefont {Swinney},\ and\
  \citenamefont {Zhang}}]{2012SwinEtAl}%
  \BibitemOpen
  \bibfield  {author} {\bibinfo {author} {\bibfnamefont {X.}~\bibnamefont
  {Chen}}, \bibinfo {author} {\bibfnamefont {X.}~\bibnamefont {Dong}}, \bibinfo
  {author} {\bibfnamefont {A.}~\bibnamefont {Be'er}}, \bibinfo {author}
  {\bibfnamefont {H.~L.}\ \bibnamefont {Swinney}}, \ and\ \bibinfo {author}
  {\bibfnamefont {H.~P.}\ \bibnamefont {Zhang}},\ }\href {\doibase
  10.1103/PhysRevLett.108.148101} {\bibfield  {journal} {\bibinfo  {journal}
  {Phys. Rev. Lett.}\ }\textbf {\bibinfo {volume} {108}},\ \bibinfo {pages}
  {148101} (\bibinfo {year} {2012})}\BibitemShut {NoStop}%
\bibitem [{\citenamefont {Sokolov}\ and\ \citenamefont
  {Aranson}(2012)}]{Sokolov_PRL12}%
  \BibitemOpen
  \bibfield  {author} {\bibinfo {author} {\bibfnamefont {A.}~\bibnamefont
  {Sokolov}}\ and\ \bibinfo {author} {\bibfnamefont {I.~S.}\ \bibnamefont
  {Aranson}},\ }\href {\doibase 10.1103/PhysRevLett.109.248109} {\bibfield
  {journal} {\bibinfo  {journal} {Phys. Rev. Lett.}\ }\textbf {\bibinfo
  {volume} {109}},\ \bibinfo {pages} {248109} (\bibinfo {year}
  {2012})}\BibitemShut {NoStop}%
\bibitem [{\citenamefont {Saintillan}\ and\ \citenamefont
  {Shelley}(2008)}]{2008Saint_Shelley}%
  \BibitemOpen
  \bibfield  {author} {\bibinfo {author} {\bibfnamefont {D.}~\bibnamefont
  {Saintillan}}\ and\ \bibinfo {author} {\bibfnamefont {M.~J.}\ \bibnamefont
  {Shelley}},\ }\href {\doibase 10.1063/1.3041776} {\bibfield  {journal}
  {\bibinfo  {journal} {Phys. Fluids}\ }\textbf {\bibinfo {volume} {20}},\
  \bibinfo {pages} {123304} (\bibinfo {year} {2008})}\BibitemShut {NoStop}%
\bibitem [{\citenamefont {Wensink}\ \emph {et~al.}(2012)\citenamefont
  {Wensink}, \citenamefont {Dunkel}, \citenamefont {Heidenreich}, \citenamefont
  {Drescher}, \citenamefont {Goldstein}, \citenamefont {L\"owen},\ and\
  \citenamefont {Yeomans}}]{PNAS}%
  \BibitemOpen
  \bibfield  {author} {\bibinfo {author} {\bibfnamefont {H.~H.}\ \bibnamefont
  {Wensink}}, \bibinfo {author} {\bibfnamefont {J.}~\bibnamefont {Dunkel}},
  \bibinfo {author} {\bibfnamefont {S.}~\bibnamefont {Heidenreich}}, \bibinfo
  {author} {\bibfnamefont {K.}~\bibnamefont {Drescher}}, \bibinfo {author}
  {\bibfnamefont {R.~E.}\ \bibnamefont {Goldstein}}, \bibinfo {author}
  {\bibfnamefont {H.}~\bibnamefont {L\"owen}}, \ and\ \bibinfo {author}
  {\bibfnamefont {J.~M.}\ \bibnamefont {Yeomans}},\ }\href {\doibase
  10.1073/pnas.1202032109} {\bibfield  {journal} {\bibinfo  {journal} {Proc.
  Natl. Acad. Sci. U.S.A.}\ }\textbf {\bibinfo {volume} {109}},\ \bibinfo
  {pages} {14308} (\bibinfo {year} {2012})}\BibitemShut {NoStop}%
\bibitem [{\citenamefont {Liu}\ and\ \citenamefont {I}(2013)}]{Li_PRE}%
  \BibitemOpen
  \bibfield  {author} {\bibinfo {author} {\bibfnamefont {K.-A.}\ \bibnamefont
  {Liu}}\ and\ \bibinfo {author} {\bibfnamefont {L.}~\bibnamefont {I}},\ }\href
  {\doibase 10.1103/PhysRevE.88.033004} {\bibfield  {journal} {\bibinfo
  {journal} {Phys. Rev. E}\ }\textbf {\bibinfo {volume} {88}},\ \bibinfo
  {pages} {033004} (\bibinfo {year} {2013})}\BibitemShut {NoStop}%
\bibitem [{\citenamefont {Aranson}(2013{\natexlab{b}})}]{aranson2013aquatic}%
  \BibitemOpen
  \bibfield  {author} {\bibinfo {author} {\bibfnamefont {I.}~\bibnamefont
  {Aranson}},\ }\href@noop {} {\bibfield  {journal} {\bibinfo  {journal}
  {Physics}\ }\textbf {\bibinfo {volume} {6}},\ \bibinfo {pages} {61} (\bibinfo
  {year} {2013}{\natexlab{b}})}\BibitemShut {NoStop}%
\bibitem [{\citenamefont {Abkenar}\ \emph {et~al.}(2013)\citenamefont
  {Abkenar}, \citenamefont {Marx}, \citenamefont {Auth},\ and\ \citenamefont
  {Gompper}}]{gompper2013}%
  \BibitemOpen
  \bibfield  {author} {\bibinfo {author} {\bibfnamefont {M.}~\bibnamefont
  {Abkenar}}, \bibinfo {author} {\bibfnamefont {K.}~\bibnamefont {Marx}},
  \bibinfo {author} {\bibfnamefont {T.}~\bibnamefont {Auth}}, \ and\ \bibinfo
  {author} {\bibfnamefont {G.}~\bibnamefont {Gompper}},\ }\href {\doibase
  10.1103/PhysRevE.88.062314} {\bibfield  {journal} {\bibinfo  {journal} {Phys.
  Rev. E}\ }\textbf {\bibinfo {volume} {88}},\ \bibinfo {pages} {062314}
  (\bibinfo {year} {2013})}\BibitemShut {NoStop}%
\bibitem [{\citenamefont {Rabani}\ \emph {et~al.}(2013)\citenamefont {Rabani},
  \citenamefont {Ariel},\ and\ \citenamefont {Be'er}}]{CollectiveSpheres}%
  \BibitemOpen
  \bibfield  {author} {\bibinfo {author} {\bibfnamefont {A.}~\bibnamefont
  {Rabani}}, \bibinfo {author} {\bibfnamefont {G.}~\bibnamefont {Ariel}}, \
  and\ \bibinfo {author} {\bibfnamefont {A.}~\bibnamefont {Be'er}},\ }\href
  {\doibase 10.1371/journal.pone.0083760} {\bibfield  {journal} {\bibinfo
  {journal} {PLoS ONE}\ }\textbf {\bibinfo {volume} {8}},\ \bibinfo {pages}
  {e83760} (\bibinfo {year} {2013})}\BibitemShut {NoStop}%
\bibitem [{\citenamefont {Zhou}\ \emph {et~al.}(2014)\citenamefont {Zhou},
  \citenamefont {Sokolov}, \citenamefont {Lavrentovich},\ and\ \citenamefont
  {Aranson}}]{ISA-PNAS14}%
  \BibitemOpen
  \bibfield  {author} {\bibinfo {author} {\bibfnamefont {S.}~\bibnamefont
  {Zhou}}, \bibinfo {author} {\bibfnamefont {A.}~\bibnamefont {Sokolov}},
  \bibinfo {author} {\bibfnamefont {O.~D.}\ \bibnamefont {Lavrentovich}}, \
  and\ \bibinfo {author} {\bibfnamefont {I.~S.}\ \bibnamefont {Aranson}},\
  }\href@noop {} {\bibfield  {journal} {\bibinfo  {journal} {Proc. Natl. Acad.
  Sci. U.S.A.}\ }\textbf {\bibinfo {volume} {111}},\ \bibinfo {pages} {1265}
  (\bibinfo {year} {2014})}\BibitemShut {NoStop}%
\bibitem [{\citenamefont {Lushi}\ \emph {et~al.}(2014)\citenamefont {Lushi},
  \citenamefont {Wioland},\ and\ \citenamefont {Goldstein}}]{WiolandPNAS2014}%
  \BibitemOpen
  \bibfield  {author} {\bibinfo {author} {\bibfnamefont {E.}~\bibnamefont
  {Lushi}}, \bibinfo {author} {\bibfnamefont {H.}~\bibnamefont {Wioland}}, \
  and\ \bibinfo {author} {\bibfnamefont {R.~E.}\ \bibnamefont {Goldstein}},\
  }\href {\doibase 10.1073/pnas.1405698111} {\bibfield  {journal} {\bibinfo
  {journal} {Proc. Natl. Acad. Sci. U.S.A.}\ }\textbf {\bibinfo {volume}
  {111}},\ \bibinfo {pages} {9733} (\bibinfo {year} {2014})}\BibitemShut
  {NoStop}%
\bibitem [{\citenamefont {Wensink}\ and\ \citenamefont
  {L\"owen}(2012)}]{WensinkJPCM}%
  \BibitemOpen
  \bibfield  {author} {\bibinfo {author} {\bibfnamefont {H.~H.}\ \bibnamefont
  {Wensink}}\ and\ \bibinfo {author} {\bibfnamefont {H.}~\bibnamefont
  {L\"owen}},\ }\href {\doibase 10.1088/0953-8984/24/46/464130} {\bibfield
  {journal} {\bibinfo  {journal} {J. Phys. Condens. Matter}\ }\textbf {\bibinfo
  {volume} {24}},\ \bibinfo {pages} {464130} (\bibinfo {year}
  {2012})}\BibitemShut {NoStop}%
\bibitem [{\citenamefont {Bricard}\ \emph {et~al.}(2013)\citenamefont
  {Bricard}, \citenamefont {Caussin}, \citenamefont {Desreumaux}, \citenamefont
  {Dauchot},\ and\ \citenamefont {Bartolo}}]{ActiveRollers}%
  \BibitemOpen
  \bibfield  {author} {\bibinfo {author} {\bibfnamefont {A.}~\bibnamefont
  {Bricard}}, \bibinfo {author} {\bibfnamefont {J.-B.}\ \bibnamefont
  {Caussin}}, \bibinfo {author} {\bibfnamefont {N.}~\bibnamefont {Desreumaux}},
  \bibinfo {author} {\bibfnamefont {O.}~\bibnamefont {Dauchot}}, \ and\
  \bibinfo {author} {\bibfnamefont {D.}~\bibnamefont {Bartolo}},\ }\href
  {\doibase 10.1038/nature12673} {\bibfield  {journal} {\bibinfo  {journal}
  {Nature}\ }\textbf {\bibinfo {volume} {503}},\ \bibinfo {pages} {95}
  (\bibinfo {year} {2013})}\BibitemShut {NoStop}%
\bibitem [{\citenamefont {Ferrante}\ \emph {et~al.}(2013)\citenamefont
  {Ferrante}, \citenamefont {Turgut}, \citenamefont {Dorigo},\ and\
  \citenamefont {Huepe}}]{Huepe_PRL}%
  \BibitemOpen
  \bibfield  {author} {\bibinfo {author} {\bibfnamefont {E.}~\bibnamefont
  {Ferrante}}, \bibinfo {author} {\bibfnamefont {A.~E.}\ \bibnamefont
  {Turgut}}, \bibinfo {author} {\bibfnamefont {M.}~\bibnamefont {Dorigo}}, \
  and\ \bibinfo {author} {\bibfnamefont {C.}~\bibnamefont {Huepe}},\ }\href
  {\doibase 10.1103/PhysRevLett.111.268302} {\bibfield  {journal} {\bibinfo
  {journal} {Phys. Rev. Lett.}\ }\textbf {\bibinfo {volume} {111}},\ \bibinfo
  {pages} {268302} (\bibinfo {year} {2013})}\BibitemShut {NoStop}%
\bibitem [{\citenamefont {Vicsek}\ \emph {et~al.}(1995)\citenamefont {Vicsek},
  \citenamefont {Czir\'ok}, \citenamefont {Ben-Jacob}, \citenamefont {Cohen},\
  and\ \citenamefont {Shochet}}]{vicsek_prl}%
  \BibitemOpen
  \bibfield  {author} {\bibinfo {author} {\bibfnamefont {T.}~\bibnamefont
  {Vicsek}}, \bibinfo {author} {\bibfnamefont {A.}~\bibnamefont {Czir\'ok}},
  \bibinfo {author} {\bibfnamefont {E.}~\bibnamefont {Ben-Jacob}}, \bibinfo
  {author} {\bibfnamefont {I.}~\bibnamefont {Cohen}}, \ and\ \bibinfo {author}
  {\bibfnamefont {O.}~\bibnamefont {Shochet}},\ }\href {\doibase
  10.1103/PhysRevLett.75.1226} {\bibfield  {journal} {\bibinfo  {journal}
  {Phys. Rev. Lett.}\ }\textbf {\bibinfo {volume} {75}},\ \bibinfo {pages}
  {1226} (\bibinfo {year} {1995})}\BibitemShut {NoStop}%
\bibitem [{\citenamefont {Kudrolli}\ \emph {et~al.}(2008)\citenamefont
  {Kudrolli}, \citenamefont {Lumay}, \citenamefont {Volfson},\ and\
  \citenamefont {Tsimring}}]{08kudrolli}%
  \BibitemOpen
  \bibfield  {author} {\bibinfo {author} {\bibfnamefont {A.}~\bibnamefont
  {Kudrolli}}, \bibinfo {author} {\bibfnamefont {G.}~\bibnamefont {Lumay}},
  \bibinfo {author} {\bibfnamefont {D.}~\bibnamefont {Volfson}}, \ and\
  \bibinfo {author} {\bibfnamefont {L.~S.}\ \bibnamefont {Tsimring}},\ }\href
  {\doibase 10.1103/PhysRevLett.100.058001} {\bibfield  {journal} {\bibinfo
  {journal} {Phys. Rev. Lett.}\ }\textbf {\bibinfo {volume} {100}},\ \bibinfo
  {pages} {058001} (\bibinfo {year} {2008})}\BibitemShut {NoStop}%
\bibitem [{\citenamefont {Narayan}\ \emph {et~al.}(2007)\citenamefont
  {Narayan}, \citenamefont {Ramaswamy},\ and\ \citenamefont
  {Menon}}]{2007NaRaMe}%
  \BibitemOpen
  \bibfield  {author} {\bibinfo {author} {\bibfnamefont {V.}~\bibnamefont
  {Narayan}}, \bibinfo {author} {\bibfnamefont {S.}~\bibnamefont {Ramaswamy}},
  \ and\ \bibinfo {author} {\bibfnamefont {N.}~\bibnamefont {Menon}},\ }\href
  {\doibase 10.1126/science.1140414} {\bibfield  {journal} {\bibinfo  {journal}
  {Science}\ }\textbf {\bibinfo {volume} {317}},\ \bibinfo {pages} {5834}
  (\bibinfo {year} {2007})}\BibitemShut {NoStop}%
\bibitem [{\citenamefont {Wang}\ \emph {et~al.}(2013)\citenamefont {Wang},
  \citenamefont {Duan}, \citenamefont {Sen},\ and\ \citenamefont
  {Mallouk}}]{SenPNAS13}%
  \BibitemOpen
  \bibfield  {author} {\bibinfo {author} {\bibfnamefont {W.}~\bibnamefont
  {Wang}}, \bibinfo {author} {\bibfnamefont {W.}~\bibnamefont {Duan}}, \bibinfo
  {author} {\bibfnamefont {A.}~\bibnamefont {Sen}}, \ and\ \bibinfo {author}
  {\bibfnamefont {T.~E.}\ \bibnamefont {Mallouk}},\ }\href@noop {} {\bibfield
  {journal} {\bibinfo  {journal} {Proc. Natl. Acad. Sci. U.S.A.}\ }\textbf
  {\bibinfo {volume} {110}},\ \bibinfo {pages} {17744} (\bibinfo {year}
  {2013})}\BibitemShut {NoStop}%
\bibitem [{\citenamefont {Aranson}\ \emph {et~al.}(2007)\citenamefont
  {Aranson}, \citenamefont {Sokolov}, \citenamefont {Kessler},\ and\
  \citenamefont {Goldstein}}]{aranson2007}%
  \BibitemOpen
  \bibfield  {author} {\bibinfo {author} {\bibfnamefont {I.~S.}\ \bibnamefont
  {Aranson}}, \bibinfo {author} {\bibfnamefont {A.}~\bibnamefont {Sokolov}},
  \bibinfo {author} {\bibfnamefont {J.~O.}\ \bibnamefont {Kessler}}, \ and\
  \bibinfo {author} {\bibfnamefont {R.~E.}\ \bibnamefont {Goldstein}},\
  }\href@noop {} {\bibfield  {journal} {\bibinfo  {journal} {Phys. Rev. E}\
  }\textbf {\bibinfo {volume} {75}},\ \bibinfo {pages} {040901} (\bibinfo
  {year} {2007})}\BibitemShut {NoStop}%
\bibitem [{\citenamefont {Drescher}\ \emph {et~al.}(2011)\citenamefont
  {Drescher}, \citenamefont {Dunkel}, \citenamefont {Cisneros}, \citenamefont
  {Ganguly},\ and\ \citenamefont {Goldstein}}]{drescher2011fluid}%
  \BibitemOpen
  \bibfield  {author} {\bibinfo {author} {\bibfnamefont {K.}~\bibnamefont
  {Drescher}}, \bibinfo {author} {\bibfnamefont {J.}~\bibnamefont {Dunkel}},
  \bibinfo {author} {\bibfnamefont {L.~H.}\ \bibnamefont {Cisneros}}, \bibinfo
  {author} {\bibfnamefont {S.}~\bibnamefont {Ganguly}}, \ and\ \bibinfo
  {author} {\bibfnamefont {R.~E.}\ \bibnamefont {Goldstein}},\ }\href@noop {}
  {\bibfield  {journal} {\bibinfo  {journal} {Proc. Natl. Acad. Sci. U.S.A.}\
  }\textbf {\bibinfo {volume} {108}},\ \bibinfo {pages} {10940} (\bibinfo
  {year} {2011})}\BibitemShut {NoStop}%
\bibitem [{\citenamefont {Denissenko}\ \emph {et~al.}(2012)\citenamefont
  {Denissenko}, \citenamefont {Kantsler}, \citenamefont {Smith},\ and\
  \citenamefont {Kirkman-Brown}}]{DenissenkoPNAS}%
  \BibitemOpen
  \bibfield  {author} {\bibinfo {author} {\bibfnamefont {P.}~\bibnamefont
  {Denissenko}}, \bibinfo {author} {\bibfnamefont {V.}~\bibnamefont
  {Kantsler}}, \bibinfo {author} {\bibfnamefont {D.~J.}\ \bibnamefont {Smith}},
  \ and\ \bibinfo {author} {\bibfnamefont {J.}~\bibnamefont {Kirkman-Brown}},\
  }\href {\doibase 10.1073/pnas.1202934109} {\bibfield  {journal} {\bibinfo
  {journal} {Proc. Natl. Acad. Sci. U.S.A.}\ }\textbf {\bibinfo {volume}
  {109}},\ \bibinfo {pages} {8007} (\bibinfo {year} {2012})}\BibitemShut
  {NoStop}%
\bibitem [{\citenamefont {Guidobaldi}\ \emph {et~al.}(2014)\citenamefont
  {Guidobaldi}, \citenamefont {Jeyaram}, \citenamefont {Berdakin},
  \citenamefont {Moshchalkov}, \citenamefont {Condat}, \citenamefont {Marconi},
  \citenamefont {Giojalas},\ and\ \citenamefont {Silhanek}}]{TrappingSperms}%
  \BibitemOpen
  \bibfield  {author} {\bibinfo {author} {\bibfnamefont {A.}~\bibnamefont
  {Guidobaldi}}, \bibinfo {author} {\bibfnamefont {Y.}~\bibnamefont {Jeyaram}},
  \bibinfo {author} {\bibfnamefont {I.}~\bibnamefont {Berdakin}}, \bibinfo
  {author} {\bibfnamefont {V.~V.}\ \bibnamefont {Moshchalkov}}, \bibinfo
  {author} {\bibfnamefont {C.~A.}\ \bibnamefont {Condat}}, \bibinfo {author}
  {\bibfnamefont {V.~I.}\ \bibnamefont {Marconi}}, \bibinfo {author}
  {\bibfnamefont {L.}~\bibnamefont {Giojalas}}, \ and\ \bibinfo {author}
  {\bibfnamefont {A.~V.}\ \bibnamefont {Silhanek}},\ }\href {\doibase
  10.1103/PhysRevE.89.032720} {\bibfield  {journal} {\bibinfo  {journal} {Phys.
  Rev. E}\ }\textbf {\bibinfo {volume} {89}},\ \bibinfo {pages} {032720}
  (\bibinfo {year} {2014})}\BibitemShut {NoStop}%
\bibitem [{\citenamefont {Wensink}\ and\ \citenamefont
  {L\"owen}(2008)}]{Wensink2008}%
  \BibitemOpen
  \bibfield  {author} {\bibinfo {author} {\bibfnamefont {H.~H.}\ \bibnamefont
  {Wensink}}\ and\ \bibinfo {author} {\bibfnamefont {H.}~\bibnamefont
  {L\"owen}},\ }\href {\doibase 10.1103/PhysRevE.78.031409} {\bibfield
  {journal} {\bibinfo  {journal} {Phys. Rev. E}\ }\textbf {\bibinfo {volume}
  {78}},\ \bibinfo {pages} {031409} (\bibinfo {year} {2008})}\BibitemShut
  {NoStop}%
\bibitem [{\citenamefont {Menzel}(2013)}]{MenzelLaningConf}%
  \BibitemOpen
  \bibfield  {author} {\bibinfo {author} {\bibfnamefont {A.~M.}\ \bibnamefont
  {Menzel}},\ }\href {http://stacks.iop.org/0953-8984/25/i=50/a=505103}
  {\bibfield  {journal} {\bibinfo  {journal} {J. Phys. Condens. Matter}\
  }\textbf {\bibinfo {volume} {25}},\ \bibinfo {pages} {505103} (\bibinfo
  {year} {2013})}\BibitemShut {NoStop}%
\bibitem [{\citenamefont {Lee}(2013)}]{Lee13Wall}%
  \BibitemOpen
  \bibfield  {author} {\bibinfo {author} {\bibfnamefont {C.~F.}\ \bibnamefont
  {Lee}},\ }\href@noop {} {\bibfield  {journal} {\bibinfo  {journal} {New J.
  Phys.}\ }\textbf {\bibinfo {volume} {15}},\ \bibinfo {pages} {055007}
  (\bibinfo {year} {2013})}\BibitemShut {NoStop}%
\bibitem [{\citenamefont {Elgeti}\ and\ \citenamefont
  {Gompper}(2013)}]{ElgetiGompper13}%
  \BibitemOpen
  \bibfield  {author} {\bibinfo {author} {\bibfnamefont {J.}~\bibnamefont
  {Elgeti}}\ and\ \bibinfo {author} {\bibfnamefont {G.}~\bibnamefont
  {Gompper}},\ }\href@noop {} {\bibfield  {journal} {\bibinfo  {journal}
  {Europhys. Lett.}\ }\textbf {\bibinfo {volume} {101}},\ \bibinfo {pages}
  {48003} (\bibinfo {year} {2013})}\BibitemShut {NoStop}%
\bibitem [{\citenamefont {Fily}\ \emph {et~al.}(2014)\citenamefont {Fily},
  \citenamefont {Baskaran},\ and\ \citenamefont {Hagan}}]{HaganConfinement}%
  \BibitemOpen
  \bibfield  {author} {\bibinfo {author} {\bibfnamefont {Y.}~\bibnamefont
  {Fily}}, \bibinfo {author} {\bibfnamefont {A.}~\bibnamefont {Baskaran}}, \
  and\ \bibinfo {author} {\bibfnamefont {M.~F.}\ \bibnamefont {Hagan}},\
  }\href@noop {} {\bibfield  {journal} {\bibinfo  {journal} {arXiv:1402.5583}\
  } (\bibinfo {year} {2014})}\BibitemShut {NoStop}%
\bibitem [{\citenamefont {Wan}\ \emph {et~al.}(2008)\citenamefont {Wan},
  \citenamefont {Olson~Reichhardt}, \citenamefont {Nussinov},\ and\
  \citenamefont {Reichhardt}}]{Reichhardt_PRL2008}%
  \BibitemOpen
  \bibfield  {author} {\bibinfo {author} {\bibfnamefont {M.~B.}\ \bibnamefont
  {Wan}}, \bibinfo {author} {\bibfnamefont {C.~J.}\ \bibnamefont
  {Olson~Reichhardt}}, \bibinfo {author} {\bibfnamefont {Z.}~\bibnamefont
  {Nussinov}}, \ and\ \bibinfo {author} {\bibfnamefont {C.}~\bibnamefont
  {Reichhardt}},\ }\href {\doibase 10.1103/PhysRevLett.101.018102} {\bibfield
  {journal} {\bibinfo  {journal} {Phys. Rev. Lett.}\ }\textbf {\bibinfo
  {volume} {101}},\ \bibinfo {pages} {018102} (\bibinfo {year}
  {2008})}\BibitemShut {NoStop}%
\bibitem [{\citenamefont {Tailleur}\ and\ \citenamefont
  {Cates}(2009)}]{CatesTrapping}%
  \BibitemOpen
  \bibfield  {author} {\bibinfo {author} {\bibfnamefont {J.}~\bibnamefont
  {Tailleur}}\ and\ \bibinfo {author} {\bibfnamefont {M.~E.}\ \bibnamefont
  {Cates}},\ }\href@noop {} {\bibfield  {journal} {\bibinfo  {journal}
  {Europhys. Lett.}\ }\textbf {\bibinfo {volume} {86}},\ \bibinfo {pages}
  {60002} (\bibinfo {year} {2009})}\BibitemShut {NoStop}%
\bibitem [{\citenamefont {Pototsky}\ \emph {et~al.}(2013)\citenamefont
  {Pototsky}, \citenamefont {Hahn},\ and\ \citenamefont
  {Stark}}]{StarkPRE13Rectification}%
  \BibitemOpen
  \bibfield  {author} {\bibinfo {author} {\bibfnamefont {A.}~\bibnamefont
  {Pototsky}}, \bibinfo {author} {\bibfnamefont {A.~M.}\ \bibnamefont {Hahn}},
  \ and\ \bibinfo {author} {\bibfnamefont {H.}~\bibnamefont {Stark}},\ }\href
  {\doibase 10.1103/PhysRevE.87.042124} {\bibfield  {journal} {\bibinfo
  {journal} {Phys. Rev. E}\ }\textbf {\bibinfo {volume} {87}},\ \bibinfo
  {pages} {042124} (\bibinfo {year} {2013})}\BibitemShut {NoStop}%
\bibitem [{\citenamefont {Potiguar}\ \emph {et~al.}(2014)\citenamefont
  {Potiguar}, \citenamefont {Farias},\ and\ \citenamefont
  {Ferreira}}]{RectifyBrazil}%
  \BibitemOpen
  \bibfield  {author} {\bibinfo {author} {\bibfnamefont {F.~Q.}\ \bibnamefont
  {Potiguar}}, \bibinfo {author} {\bibfnamefont {G.~A.}\ \bibnamefont
  {Farias}}, \ and\ \bibinfo {author} {\bibfnamefont {W.~P.}\ \bibnamefont
  {Ferreira}},\ }\href {\doibase 10.1103/PhysRevE.90.012307} {\bibfield
  {journal} {\bibinfo  {journal} {Phys. Rev. E}\ }\textbf {\bibinfo {volume}
  {90}},\ \bibinfo {pages} {012307} (\bibinfo {year} {2014})}\BibitemShut
  {NoStop}%
\bibitem [{\citenamefont {Galajda}\ \emph {et~al.}(2007)\citenamefont
  {Galajda}, \citenamefont {Keymer}, \citenamefont {Chaikin},\ and\
  \citenamefont {Austin}}]{Chaikin2007}%
  \BibitemOpen
  \bibfield  {author} {\bibinfo {author} {\bibfnamefont {P.}~\bibnamefont
  {Galajda}}, \bibinfo {author} {\bibfnamefont {J.}~\bibnamefont {Keymer}},
  \bibinfo {author} {\bibfnamefont {P.}~\bibnamefont {Chaikin}}, \ and\
  \bibinfo {author} {\bibfnamefont {R.}~\bibnamefont {Austin}},\ }\href@noop {}
  {\bibfield  {journal} {\bibinfo  {journal} {J. Bacteriol.}\ }\textbf
  {\bibinfo {volume} {189}},\ \bibinfo {pages} {8704} (\bibinfo {year}
  {2007})}\BibitemShut {NoStop}%
\bibitem [{\citenamefont {Hulme}\ \emph {et~al.}(2008)\citenamefont {Hulme},
  \citenamefont {DiLuzio}, \citenamefont {Shevkoplyas}, \citenamefont {Turner},
  \citenamefont {Mayer}, \citenamefont {Berg},\ and\ \citenamefont
  {Whitesides}}]{hulme}%
  \BibitemOpen
  \bibfield  {author} {\bibinfo {author} {\bibfnamefont {S.~E.}\ \bibnamefont
  {Hulme}}, \bibinfo {author} {\bibfnamefont {W.~R.}\ \bibnamefont {DiLuzio}},
  \bibinfo {author} {\bibfnamefont {S.~S.}\ \bibnamefont {Shevkoplyas}},
  \bibinfo {author} {\bibfnamefont {L.}~\bibnamefont {Turner}}, \bibinfo
  {author} {\bibfnamefont {M.}~\bibnamefont {Mayer}}, \bibinfo {author}
  {\bibfnamefont {H.~C.}\ \bibnamefont {Berg}}, \ and\ \bibinfo {author}
  {\bibfnamefont {G.~M.}\ \bibnamefont {Whitesides}},\ }\href@noop {}
  {\bibfield  {journal} {\bibinfo  {journal} {Lab Chip}\ }\textbf {\bibinfo
  {volume} {8}},\ \bibinfo {pages} {1888} (\bibinfo {year} {2008})}\BibitemShut
  {NoStop}%
\bibitem [{\citenamefont {Drocco}\ \emph {et~al.}(2012)\citenamefont {Drocco},
  \citenamefont {Olson~Reichhardt},\ and\ \citenamefont
  {Reichhardt}}]{Reichhardt}%
  \BibitemOpen
  \bibfield  {author} {\bibinfo {author} {\bibfnamefont {J.~A.}\ \bibnamefont
  {Drocco}}, \bibinfo {author} {\bibfnamefont {C.~J.}\ \bibnamefont
  {Olson~Reichhardt}}, \ and\ \bibinfo {author} {\bibfnamefont
  {C.}~\bibnamefont {Reichhardt}},\ }\href {\doibase
  10.1103/PhysRevE.85.056102} {\bibfield  {journal} {\bibinfo  {journal} {Phys.
  Rev. E}\ }\textbf {\bibinfo {volume} {85}},\ \bibinfo {pages} {056102}
  (\bibinfo {year} {2012})}\BibitemShut {NoStop}%
\bibitem [{\citenamefont {Berdakin}\ \emph {et~al.}(2013)\citenamefont
  {Berdakin}, \citenamefont {Jeyaram}, \citenamefont {Moshchalkov},
  \citenamefont {Venken}, \citenamefont {Dierckx}, \citenamefont
  {Vanderleyden}, \citenamefont {Silhanek}, \citenamefont {Condat},\ and\
  \citenamefont {Marconi}}]{MarconiPRE}%
  \BibitemOpen
  \bibfield  {author} {\bibinfo {author} {\bibfnamefont {I.}~\bibnamefont
  {Berdakin}}, \bibinfo {author} {\bibfnamefont {Y.}~\bibnamefont {Jeyaram}},
  \bibinfo {author} {\bibfnamefont {V.~V.}\ \bibnamefont {Moshchalkov}},
  \bibinfo {author} {\bibfnamefont {L.}~\bibnamefont {Venken}}, \bibinfo
  {author} {\bibfnamefont {S.}~\bibnamefont {Dierckx}}, \bibinfo {author}
  {\bibfnamefont {S.~J.}\ \bibnamefont {Vanderleyden}}, \bibinfo {author}
  {\bibfnamefont {A.~V.}\ \bibnamefont {Silhanek}}, \bibinfo {author}
  {\bibfnamefont {C.~A.}\ \bibnamefont {Condat}}, \ and\ \bibinfo {author}
  {\bibfnamefont {V.~I.}\ \bibnamefont {Marconi}},\ }\href {\doibase
  10.1103/PhysRevE.87.052702} {\bibfield  {journal} {\bibinfo  {journal} {Phys.
  Rev. E}\ }\textbf {\bibinfo {volume} {87}},\ \bibinfo {pages} {052702}
  (\bibinfo {year} {2013})}\BibitemShut {NoStop}%
\bibitem [{\citenamefont {Costanzo}\ \emph {et~al.}(2014)\citenamefont
  {Costanzo}, \citenamefont {Elgeti}, \citenamefont {Auth}, \citenamefont
  {Gompper},\ and\ \citenamefont {Ripoll}}]{JulichSortingMicrochannel}%
  \BibitemOpen
  \bibfield  {author} {\bibinfo {author} {\bibfnamefont {A.}~\bibnamefont
  {Costanzo}}, \bibinfo {author} {\bibfnamefont {J.}~\bibnamefont {Elgeti}},
  \bibinfo {author} {\bibfnamefont {T.}~\bibnamefont {Auth}}, \bibinfo {author}
  {\bibfnamefont {G.}~\bibnamefont {Gompper}}, \ and\ \bibinfo {author}
  {\bibfnamefont {M.}~\bibnamefont {Ripoll}},\ }\href@noop {} {\bibfield
  {journal} {\bibinfo  {journal} {Europhys. Lett.}\ }\textbf {\bibinfo {volume}
  {107}},\ \bibinfo {pages} {36003} (\bibinfo {year} {2014})}\BibitemShut
  {NoStop}%
\bibitem [{\citenamefont {Kaiser}\ \emph {et~al.}(2012)\citenamefont {Kaiser},
  \citenamefont {Wensink},\ and\ \citenamefont {L\"owen}}]{Kaiser_PRL}%
  \BibitemOpen
  \bibfield  {author} {\bibinfo {author} {\bibfnamefont {A.}~\bibnamefont
  {Kaiser}}, \bibinfo {author} {\bibfnamefont {H.~H.}\ \bibnamefont {Wensink}},
  \ and\ \bibinfo {author} {\bibfnamefont {H.}~\bibnamefont {L\"owen}},\ }\href
  {\doibase 10.1103/PhysRevLett.108.268307} {\bibfield  {journal} {\bibinfo
  {journal} {Phys. Rev. Lett.}\ }\textbf {\bibinfo {volume} {108}},\ \bibinfo
  {pages} {268307} (\bibinfo {year} {2012})}\BibitemShut {NoStop}%
\bibitem [{\citenamefont {Restrepo-P{\'e}rez}\ \emph
  {et~al.}(2014)\citenamefont {Restrepo-P{\'e}rez}, \citenamefont {Soler},
  \citenamefont {Mart{\'\i}nez-Cisneros}, \citenamefont {Sanchez},\ and\
  \citenamefont {Schmidt}}]{TrappingOGS}%
  \BibitemOpen
  \bibfield  {author} {\bibinfo {author} {\bibfnamefont {L.}~\bibnamefont
  {Restrepo-P{\'e}rez}}, \bibinfo {author} {\bibfnamefont {L.}~\bibnamefont
  {Soler}}, \bibinfo {author} {\bibfnamefont {C.~S.}\ \bibnamefont
  {Mart{\'\i}nez-Cisneros}}, \bibinfo {author} {\bibfnamefont {S.}~\bibnamefont
  {Sanchez}}, \ and\ \bibinfo {author} {\bibfnamefont {O.~G.}\ \bibnamefont
  {Schmidt}},\ }\href@noop {} {\bibfield  {journal} {\bibinfo  {journal} {Lab
  Chip}\ }\textbf {\bibinfo {volume} {14}},\ \bibinfo {pages} {1515} (\bibinfo
  {year} {2014})}\BibitemShut {NoStop}%
\bibitem [{\citenamefont {Vagias}\ \emph {et~al.}(2013)\citenamefont {Vagias},
  \citenamefont {Raccis}, \citenamefont {Koynov}, \citenamefont {Jonas},
  \citenamefont {Butt}, \citenamefont {Fytas}, \citenamefont
  {Ko\ifmmode~\check{s}\else \v{s}\fi{}ovan}, \citenamefont {Lenz},\ and\
  \citenamefont {Holm}}]{TracerHolmPRL}%
  \BibitemOpen
  \bibfield  {author} {\bibinfo {author} {\bibfnamefont {A.}~\bibnamefont
  {Vagias}}, \bibinfo {author} {\bibfnamefont {R.}~\bibnamefont {Raccis}},
  \bibinfo {author} {\bibfnamefont {K.}~\bibnamefont {Koynov}}, \bibinfo
  {author} {\bibfnamefont {U.}~\bibnamefont {Jonas}}, \bibinfo {author}
  {\bibfnamefont {H.-J.}\ \bibnamefont {Butt}}, \bibinfo {author}
  {\bibfnamefont {G.}~\bibnamefont {Fytas}}, \bibinfo {author} {\bibfnamefont
  {P.}~\bibnamefont {Ko\ifmmode~\check{s}\else \v{s}\fi{}ovan}}, \bibinfo
  {author} {\bibfnamefont {O.}~\bibnamefont {Lenz}}, \ and\ \bibinfo {author}
  {\bibfnamefont {C.}~\bibnamefont {Holm}},\ }\href {\doibase
  10.1103/PhysRevLett.111.088301} {\bibfield  {journal} {\bibinfo  {journal}
  {Phys. Rev. Lett.}\ }\textbf {\bibinfo {volume} {111}},\ \bibinfo {pages}
  {088301} (\bibinfo {year} {2013})}\BibitemShut {NoStop}%
\bibitem [{\citenamefont {Leptos}\ \emph {et~al.}(2009)\citenamefont {Leptos},
  \citenamefont {Guasto}, \citenamefont {Gollub}, \citenamefont {Pesci},\ and\
  \citenamefont {Goldstein}}]{TracerGoldstein}%
  \BibitemOpen
  \bibfield  {author} {\bibinfo {author} {\bibfnamefont {K.~C.}\ \bibnamefont
  {Leptos}}, \bibinfo {author} {\bibfnamefont {J.~S.}\ \bibnamefont {Guasto}},
  \bibinfo {author} {\bibfnamefont {J.~P.}\ \bibnamefont {Gollub}}, \bibinfo
  {author} {\bibfnamefont {A.~I.}\ \bibnamefont {Pesci}}, \ and\ \bibinfo
  {author} {\bibfnamefont {R.~E.}\ \bibnamefont {Goldstein}},\ }\href {\doibase
  10.1103/PhysRevLett.103.198103} {\bibfield  {journal} {\bibinfo  {journal}
  {Phys. Rev. Lett.}\ }\textbf {\bibinfo {volume} {103}},\ \bibinfo {pages}
  {198103} (\bibinfo {year} {2009})}\BibitemShut {NoStop}%
\bibitem [{\citenamefont {Mi\~no}\ \emph {et~al.}(2013)\citenamefont {Mi\~no},
  \citenamefont {Dunstan}, \citenamefont {Rousselet}, \citenamefont {Clement},\
  and\ \citenamefont {Soto}}]{TracerClement}%
  \BibitemOpen
  \bibfield  {author} {\bibinfo {author} {\bibfnamefont {G.}~\bibnamefont
  {Mi\~no}}, \bibinfo {author} {\bibfnamefont {J.}~\bibnamefont {Dunstan}},
  \bibinfo {author} {\bibfnamefont {A.}~\bibnamefont {Rousselet}}, \bibinfo
  {author} {\bibfnamefont {E.}~\bibnamefont {Clement}}, \ and\ \bibinfo
  {author} {\bibfnamefont {R.}~\bibnamefont {Soto}},\ }\href {\doibase
  10.1017/jfm.2013.304} {\bibfield  {journal} {\bibinfo  {journal} {J. Fluid
  Mech.}\ }\textbf {\bibinfo {volume} {729}},\ \bibinfo {pages} {423} (\bibinfo
  {year} {2013})}\BibitemShut {NoStop}%
\bibitem [{\citenamefont {Wu}\ and\ \citenamefont
  {Libchaber}(2000)}]{TracerDiffusion2000}%
  \BibitemOpen
  \bibfield  {author} {\bibinfo {author} {\bibfnamefont {X.-L.}\ \bibnamefont
  {Wu}}\ and\ \bibinfo {author} {\bibfnamefont {A.}~\bibnamefont {Libchaber}},\
  }\href {\doibase 10.1103/PhysRevLett.84.3017} {\bibfield  {journal} {\bibinfo
   {journal} {Phys. Rev. Lett.}\ }\textbf {\bibinfo {volume} {84}},\ \bibinfo
  {pages} {3017} (\bibinfo {year} {2000})}\BibitemShut {NoStop}%
\bibitem [{\citenamefont {Mallory}\ \emph {et~al.}(2014)\citenamefont
  {Mallory}, \citenamefont {Valeriani},\ and\ \citenamefont
  {Cacciuto}}]{MalloryArxiv}%
  \BibitemOpen
  \bibfield  {author} {\bibinfo {author} {\bibfnamefont {S.~A.}\ \bibnamefont
  {Mallory}}, \bibinfo {author} {\bibfnamefont {C.}~\bibnamefont {Valeriani}},
  \ and\ \bibinfo {author} {\bibfnamefont {A.}~\bibnamefont {Cacciuto}},\
  }\href@noop {} {\bibfield  {journal} {\bibinfo  {journal} {arXiv:1407.3418}\
  } (\bibinfo {year} {2014})}\BibitemShut {NoStop}%
\bibitem [{\citenamefont {Kaiser}\ and\ \citenamefont
  {L\"owen}(2014)}]{KaiserPolymer}%
  \BibitemOpen
  \bibfield  {author} {\bibinfo {author} {\bibfnamefont {A.}~\bibnamefont
  {Kaiser}}\ and\ \bibinfo {author} {\bibfnamefont {H.}~\bibnamefont
  {L\"owen}},\ }\href {\doibase 10.1103/PhysRevLett.112.158101} {\bibfield
  {journal} {\bibinfo  {journal} {J. Chem. Phys.}\ }\textbf {\bibinfo {volume}
  {141}},\ \bibinfo {pages} {044903} (\bibinfo {year} {2014})}\BibitemShut
  {NoStop}%
\bibitem [{\citenamefont {ten Hagen}\ \emph {et~al.}(2011)\citenamefont {ten
  Hagen}, \citenamefont {van Teeffelen},\ and\ \citenamefont
  {L\"owen}}]{BtH2011}%
  \BibitemOpen
  \bibfield  {author} {\bibinfo {author} {\bibfnamefont {B.}~\bibnamefont {ten
  Hagen}}, \bibinfo {author} {\bibfnamefont {S.}~\bibnamefont {van Teeffelen}},
  \ and\ \bibinfo {author} {\bibfnamefont {H.}~\bibnamefont {L\"owen}},\ }\href
  {\doibase 10.1088/0953-8984/23/19/194119} {\bibfield  {journal} {\bibinfo
  {journal} {J. Phys. Condens. Matter}\ }\textbf {\bibinfo {volume} {23}},\
  \bibinfo {pages} {194119} (\bibinfo {year} {2011})}\BibitemShut {NoStop}%
\bibitem [{\citenamefont {Zheng}\ \emph {et~al.}(2013)\citenamefont {Zheng},
  \citenamefont {ten Hagen}, \citenamefont {Kaiser}, \citenamefont {Wu},
  \citenamefont {Cui}, \citenamefont {Silber-Li},\ and\ \citenamefont
  {L\"owen}}]{Kaiser_PRE2013Janus}%
  \BibitemOpen
  \bibfield  {author} {\bibinfo {author} {\bibfnamefont {X.}~\bibnamefont
  {Zheng}}, \bibinfo {author} {\bibfnamefont {B.}~\bibnamefont {ten Hagen}},
  \bibinfo {author} {\bibfnamefont {A.}~\bibnamefont {Kaiser}}, \bibinfo
  {author} {\bibfnamefont {M.}~\bibnamefont {Wu}}, \bibinfo {author}
  {\bibfnamefont {H.}~\bibnamefont {Cui}}, \bibinfo {author} {\bibfnamefont
  {Z.}~\bibnamefont {Silber-Li}}, \ and\ \bibinfo {author} {\bibfnamefont
  {H.}~\bibnamefont {L\"owen}},\ }\href {\doibase 10.1103/PhysRevE.88.032304}
  {\bibfield  {journal} {\bibinfo  {journal} {Phys. Rev. E}\ }\textbf {\bibinfo
  {volume} {88}},\ \bibinfo {pages} {032304} (\bibinfo {year}
  {2013})}\BibitemShut {NoStop}%
\bibitem [{\citenamefont {Hess}(2011)}]{HessReview}%
  \BibitemOpen
  \bibfield  {author} {\bibinfo {author} {\bibfnamefont {H.}~\bibnamefont
  {Hess}},\ }\href {\doibase 10.1146/annurev-bioeng-071910-124644} {\bibfield
  {journal} {\bibinfo  {journal} {Annu. Rev. Biomed. Eng.}\ }\textbf {\bibinfo
  {volume} {13}},\ \bibinfo {pages} {429} (\bibinfo {year} {2011})}\BibitemShut
  {NoStop}%
\bibitem [{\citenamefont {Angelani}\ \emph {et~al.}(2009)\citenamefont
  {Angelani}, \citenamefont {DiLeonardo},\ and\ \citenamefont
  {Ruocco}}]{DiLeonardoPRL}%
  \BibitemOpen
  \bibfield  {author} {\bibinfo {author} {\bibfnamefont {L.}~\bibnamefont
  {Angelani}}, \bibinfo {author} {\bibfnamefont {R.}~\bibnamefont
  {DiLeonardo}}, \ and\ \bibinfo {author} {\bibfnamefont {G.}~\bibnamefont
  {Ruocco}},\ }\href@noop {} {\bibfield  {journal} {\bibinfo  {journal} {Phys.
  Rev. Lett.}\ }\textbf {\bibinfo {volume} {102}},\ \bibinfo {pages} {048104}
  (\bibinfo {year} {2009})}\BibitemShut {NoStop}%
\bibitem [{\citenamefont {Sokolov}\ \emph {et~al.}(2010)\citenamefont
  {Sokolov}, \citenamefont {Apodaca}, \citenamefont {Grzyboski},\ and\
  \citenamefont {Aranson}}]{SokolovPNAS}%
  \BibitemOpen
  \bibfield  {author} {\bibinfo {author} {\bibfnamefont {A.}~\bibnamefont
  {Sokolov}}, \bibinfo {author} {\bibfnamefont {M.~M.}\ \bibnamefont
  {Apodaca}}, \bibinfo {author} {\bibfnamefont {B.~A.}\ \bibnamefont
  {Grzyboski}}, \ and\ \bibinfo {author} {\bibfnamefont {I.~S.}\ \bibnamefont
  {Aranson}},\ }\href@noop {} {\bibfield  {journal} {\bibinfo  {journal} {Proc.
  Natl. Acad. Sci. U.S.A.}\ }\textbf {\bibinfo {volume} {107}},\ \bibinfo
  {pages} {969} (\bibinfo {year} {2010})}\BibitemShut {NoStop}%
\bibitem [{\citenamefont {DiLeonardo}\ \emph {et~al.}(2010)\citenamefont
  {DiLeonardo}, \citenamefont {Angelani}, \citenamefont {DellArciprete},
  \citenamefont {Ruocco}, \citenamefont {Iebba}, \citenamefont {Schippa},
  \citenamefont {Conte}, \citenamefont {Mecarini}, \citenamefont {Angelis},\
  and\ \citenamefont {Fabrizio}}]{LeonardoPNAS}%
  \BibitemOpen
  \bibfield  {author} {\bibinfo {author} {\bibfnamefont {R.}~\bibnamefont
  {DiLeonardo}}, \bibinfo {author} {\bibfnamefont {L.}~\bibnamefont
  {Angelani}}, \bibinfo {author} {\bibfnamefont {D.}~\bibnamefont
  {DellArciprete}}, \bibinfo {author} {\bibfnamefont {G.}~\bibnamefont
  {Ruocco}}, \bibinfo {author} {\bibfnamefont {V.}~\bibnamefont {Iebba}},
  \bibinfo {author} {\bibfnamefont {S.}~\bibnamefont {Schippa}}, \bibinfo
  {author} {\bibfnamefont {M.~P.}\ \bibnamefont {Conte}}, \bibinfo {author}
  {\bibfnamefont {F.}~\bibnamefont {Mecarini}}, \bibinfo {author}
  {\bibfnamefont {F.~D.}\ \bibnamefont {Angelis}}, \ and\ \bibinfo {author}
  {\bibfnamefont {E.~D.}\ \bibnamefont {Fabrizio}},\ }\href@noop {} {\bibfield
  {journal} {\bibinfo  {journal} {Proc. Natl. Acad. Sci. U.S.A.}\ }\textbf
  {\bibinfo {volume} {107}},\ \bibinfo {pages} {9541} (\bibinfo {year}
  {2010})}\BibitemShut {NoStop}%
\bibitem [{\citenamefont {Li}\ and\ \citenamefont {Zhang}(2013)}]{GearRobots}%
  \BibitemOpen
  \bibfield  {author} {\bibinfo {author} {\bibfnamefont {H.}~\bibnamefont
  {Li}}\ and\ \bibinfo {author} {\bibfnamefont {H.~P.}\ \bibnamefont {Zhang}},\
  }\href {http://stacks.iop.org/0295-5075/102/i=5/a=50007} {\bibfield
  {journal} {\bibinfo  {journal} {Europhys. Lett.}\ }\textbf {\bibinfo {volume}
  {102}},\ \bibinfo {pages} {50007} (\bibinfo {year} {2013})}\BibitemShut
  {NoStop}%
\bibitem [{\citenamefont {Angelani}\ and\ \citenamefont
  {Leonardo}(2010)}]{AngelaniCARGO}%
  \BibitemOpen
  \bibfield  {author} {\bibinfo {author} {\bibfnamefont {L.}~\bibnamefont
  {Angelani}}\ and\ \bibinfo {author} {\bibfnamefont {R.~D.}\ \bibnamefont
  {Leonardo}},\ }\href@noop {} {\bibfield  {journal} {\bibinfo  {journal} {New
  J. Phys.}\ }\textbf {\bibinfo {volume} {12}},\ \bibinfo {pages} {113017}
  (\bibinfo {year} {2010})}\BibitemShut {NoStop}%
\bibitem [{\citenamefont {Kaiser}\ \emph {et~al.}(2014)\citenamefont {Kaiser},
  \citenamefont {Peshkov}, \citenamefont {Sokolov}, \citenamefont {ten Hagen},
  \citenamefont {L\"owen},\ and\ \citenamefont {Aranson}}]{KaiserSokolov_2014}%
  \BibitemOpen
  \bibfield  {author} {\bibinfo {author} {\bibfnamefont {A.}~\bibnamefont
  {Kaiser}}, \bibinfo {author} {\bibfnamefont {A.}~\bibnamefont {Peshkov}},
  \bibinfo {author} {\bibfnamefont {A.}~\bibnamefont {Sokolov}}, \bibinfo
  {author} {\bibfnamefont {B.}~\bibnamefont {ten Hagen}}, \bibinfo {author}
  {\bibfnamefont {H.}~\bibnamefont {L\"owen}}, \ and\ \bibinfo {author}
  {\bibfnamefont {I.~S.}\ \bibnamefont {Aranson}},\ }\href {\doibase
  10.1103/PhysRevLett.112.158101} {\bibfield  {journal} {\bibinfo  {journal}
  {Phys. Rev. Lett.}\ }\textbf {\bibinfo {volume} {112}},\ \bibinfo {pages}
  {158101} (\bibinfo {year} {2014})}\BibitemShut {NoStop}%
\bibitem [{\citenamefont {Gachelin}\ \emph {et~al.}(2013)\citenamefont
  {Gachelin}, \citenamefont {Mi\~no}, \citenamefont {Berthet}, \citenamefont
  {Lindner}, \citenamefont {Rousselet},\ and\ \citenamefont
  {Cl\'ement}}]{Clement_PRL13}%
  \BibitemOpen
  \bibfield  {author} {\bibinfo {author} {\bibfnamefont {J.}~\bibnamefont
  {Gachelin}}, \bibinfo {author} {\bibfnamefont {G.}~\bibnamefont {Mi\~no}},
  \bibinfo {author} {\bibfnamefont {H.}~\bibnamefont {Berthet}}, \bibinfo
  {author} {\bibfnamefont {A.}~\bibnamefont {Lindner}}, \bibinfo {author}
  {\bibfnamefont {A.}~\bibnamefont {Rousselet}}, \ and\ \bibinfo {author}
  {\bibfnamefont {E.}~\bibnamefont {Cl\'ement}},\ }\href {\doibase
  10.1103/PhysRevLett.110.268103} {\bibfield  {journal} {\bibinfo  {journal}
  {Phys. Rev. Lett.}\ }\textbf {\bibinfo {volume} {110}},\ \bibinfo {pages}
  {268103} (\bibinfo {year} {2013})}\BibitemShut {NoStop}%
\bibitem [{\citenamefont {Kirchhoff}\ \emph {et~al.}(1996)\citenamefont
  {Kirchhoff}, \citenamefont {L\"owen},\ and\ \citenamefont
  {Klein}}]{Kirchhoff1996}%
  \BibitemOpen
  \bibfield  {author} {\bibinfo {author} {\bibfnamefont {T.}~\bibnamefont
  {Kirchhoff}}, \bibinfo {author} {\bibfnamefont {H.}~\bibnamefont {L\"owen}},
  \ and\ \bibinfo {author} {\bibfnamefont {R.}~\bibnamefont {Klein}},\ }\href
  {\doibase 10.1103/PhysRevE.53.5011} {\bibfield  {journal} {\bibinfo
  {journal} {Phys. Rev. E}\ }\textbf {\bibinfo {volume} {53}},\ \bibinfo
  {pages} {5011} (\bibinfo {year} {1996})}\BibitemShut {NoStop}%
\bibitem [{\citenamefont {Wensink}\ \emph {et~al.}(2014)\citenamefont
  {Wensink}, \citenamefont {Kantsler}, \citenamefont {Goldstein},\ and\
  \citenamefont {Dunkel}}]{Wensink2014}%
  \BibitemOpen
  \bibfield  {author} {\bibinfo {author} {\bibfnamefont {H.~H.}\ \bibnamefont
  {Wensink}}, \bibinfo {author} {\bibfnamefont {V.}~\bibnamefont {Kantsler}},
  \bibinfo {author} {\bibfnamefont {R.~E.}\ \bibnamefont {Goldstein}}, \ and\
  \bibinfo {author} {\bibfnamefont {J.}~\bibnamefont {Dunkel}},\ }\href
  {\doibase 10.1103/PhysRevE.89.010302} {\bibfield  {journal} {\bibinfo
  {journal} {Phys. Rev. E}\ }\textbf {\bibinfo {volume} {89}},\ \bibinfo
  {pages} {010302} (\bibinfo {year} {2014})}\BibitemShut {NoStop}%
\bibitem [{\citenamefont {Ramaswamy}(2010)}]{2010Ramaswamy}%
  \BibitemOpen
  \bibfield  {author} {\bibinfo {author} {\bibfnamefont {S.}~\bibnamefont
  {Ramaswamy}},\ }\href {\doibase 10.1146/annurev-conmatphys-070909-104101}
  {\bibfield  {journal} {\bibinfo  {journal} {Annu. Rev. Condens. Matter
  Phys.}\ }\textbf {\bibinfo {volume} {1}},\ \bibinfo {pages} {323} (\bibinfo
  {year} {2010})}\BibitemShut {NoStop}%
\bibitem [{\citenamefont {K\"ummel}\ \emph {et~al.}(2014)\citenamefont
  {K\"ummel}, \citenamefont {ten Hagen}, \citenamefont {Wittkowski},
  \citenamefont {Takagi}, \citenamefont {Buttinoni}, \citenamefont {Eichhorn},
  \citenamefont {Volpe}, \citenamefont {L\"owen},\ and\ \citenamefont
  {Bechinger}}]{BtHComment}%
  \BibitemOpen
  \bibfield  {author} {\bibinfo {author} {\bibfnamefont {F.}~\bibnamefont
  {K\"ummel}}, \bibinfo {author} {\bibfnamefont {B.}~\bibnamefont {ten Hagen}},
  \bibinfo {author} {\bibfnamefont {R.}~\bibnamefont {Wittkowski}}, \bibinfo
  {author} {\bibfnamefont {D.}~\bibnamefont {Takagi}}, \bibinfo {author}
  {\bibfnamefont {I.}~\bibnamefont {Buttinoni}}, \bibinfo {author}
  {\bibfnamefont {R.}~\bibnamefont {Eichhorn}}, \bibinfo {author}
  {\bibfnamefont {G.}~\bibnamefont {Volpe}}, \bibinfo {author} {\bibfnamefont
  {H.}~\bibnamefont {L\"owen}}, \ and\ \bibinfo {author} {\bibfnamefont
  {C.}~\bibnamefont {Bechinger}},\ }\href {\doibase
  10.1103/PhysRevLett.113.029802} {\bibfield  {journal} {\bibinfo  {journal}
  {Phys. Rev. Lett.}\ }\textbf {\bibinfo {volume} {113}},\ \bibinfo {pages}
  {029802} (\bibinfo {year} {2014})}\BibitemShut {NoStop}%
\bibitem [{\citenamefont {Tirado}\ \emph {et~al.}(1984)\citenamefont {Tirado},
  \citenamefont {Martinez},\ and\ \citenamefont {de~la Torre}}]{tirado}%
  \BibitemOpen
  \bibfield  {author} {\bibinfo {author} {\bibfnamefont {M.~M.}\ \bibnamefont
  {Tirado}}, \bibinfo {author} {\bibfnamefont {C.~L.}\ \bibnamefont
  {Martinez}}, \ and\ \bibinfo {author} {\bibfnamefont {J.~G.}\ \bibnamefont
  {de~la Torre}},\ }\href {\doibase 10.1063/1.447827} {\bibfield  {journal}
  {\bibinfo  {journal} {J. Chem. Phys.}\ }\textbf {\bibinfo {volume} {81}},\
  \bibinfo {pages} {2047} (\bibinfo {year} {1984})}\BibitemShut {NoStop}%
\bibitem [{\citenamefont {Garcia~de~la Torre}\ \emph
  {et~al.}(1994)\citenamefont {Garcia~de~la Torre}, \citenamefont {Navarro},
  \citenamefont {Lopez~Martinez}, \citenamefont {Diaz},\ and\ \citenamefont
  {Lopez~Cascales}}]{delaTorreNMDC1994}%
  \BibitemOpen
  \bibfield  {author} {\bibinfo {author} {\bibfnamefont {J.}~\bibnamefont
  {Garcia~de~la Torre}}, \bibinfo {author} {\bibfnamefont {S.}~\bibnamefont
  {Navarro}}, \bibinfo {author} {\bibfnamefont {M.~C.}\ \bibnamefont
  {Lopez~Martinez}}, \bibinfo {author} {\bibfnamefont {F.~G.}\ \bibnamefont
  {Diaz}}, \ and\ \bibinfo {author} {\bibfnamefont {J.~J.}\ \bibnamefont
  {Lopez~Cascales}},\ }\href@noop {} {\bibfield  {journal} {\bibinfo  {journal}
  {Biophys. J.}\ }\textbf {\bibinfo {volume} {67}},\ \bibinfo {pages} {530}
  (\bibinfo {year} {1994})}\BibitemShut {NoStop}%
\bibitem [{\citenamefont {Carrasco}\ and\ \citenamefont {Garcia de~la
  Torre}(1999)}]{Carrasco99}%
  \BibitemOpen
  \bibfield  {author} {\bibinfo {author} {\bibfnamefont {B.}~\bibnamefont
  {Carrasco}}\ and\ \bibinfo {author} {\bibfnamefont {J.}~\bibnamefont {Garcia
  de~la Torre}},\ }\href@noop {} {\bibfield  {journal} {\bibinfo  {journal} {J.
  Chem. Phys.}\ }\textbf {\bibinfo {volume} {111}},\ \bibinfo {pages} {4817}
  (\bibinfo {year} {1999})}\BibitemShut {NoStop}%
\bibitem [{\citenamefont {Sokolov}\ and\ \citenamefont
  {Aranson}(2009)}]{2009SoAr}%
  \BibitemOpen
  \bibfield  {author} {\bibinfo {author} {\bibfnamefont {A.}~\bibnamefont
  {Sokolov}}\ and\ \bibinfo {author} {\bibfnamefont {I.~S.}\ \bibnamefont
  {Aranson}},\ }\href {\doibase 10.1103/PhysRevLett.103.148101} {\bibfield
  {journal} {\bibinfo  {journal} {Phys. Rev. Lett.}\ }\textbf {\bibinfo
  {volume} {103}},\ \bibinfo {pages} {148101} (\bibinfo {year}
  {2009})}\BibitemShut {NoStop}%
\end{thebibliography}%
\end{document}